\documentclass[12pt,preprint]{aastex}

\slugcomment{to appear in the AJ, (October 2004 issue)}

\shorttitle{On rotation of PTTS in Associations}
\shortauthors{de la Reza \& Pinz\'on}

\begin{document}


\title{On the Rotation of Post-T Tauri Stars in Associations }


\author{Ramiro de la Reza\altaffilmark{1} and Giovanni Pinz\'on\altaffilmark{2}}
\affil{Observat\'orio Nacional, Rio de Janeiro, Brazil}






\altaffiltext{1}{delareza@on.br}
\altaffiltext{2}{gpinzon@on.br}



\begin{abstract}

Nearby associations or moving groups of post-T Tauri stars with ages between
$\sim$10 and 30 Myr are excellent objects for the study of the initial spin up
phase during the pre-main sequence evolution. An empirical approach is adopted
here for the first time with these stars to infer their rotations properties
and relations to X-ray emission. Three nearby associations with distances less
than 100 pc are considered. The TW Hya association (TWA) with an age of 8
Myr, the Beta Pictoris moving group (BPMG) with an age of 12 Myr and a combination
of Tucana and Horologium associations Tuc/HorA (30 Myr). Two low and high
rotation modes are considered for each association with stellar masses of $0.1 \leq M < 1.5$ $M_{\odot}$  and $1.5 \leq M \leq 2.6$ $M_{\odot}$ respectively. Because no observed
rotational periods are known for these stars, we use a mathematical tool to
infer representative equatorial rotation velocities $v_{0}(eq)$ from the observed distribution of projected rotational velocities ($vsini$). This is done for each mode and for each association. A spin up is found for the high rotation mode, whereas in the low
rotation mode the $v_{0}(eq)$ do not increase significantly. This insufficient increase of $v_{0}(eq)$ is probably the cause of a decrease of the total mean specific angular momentum for the low mass stars between 8 and 30 Myr. However, for the high mass stars, where a sufficient spin up is present, the specific angular momentum is practically conserved in this same time interval. A two dimensional
(mass and $vsini$) K-S statistical test yields results compatible with an spin
up scenario. By supposing that the distribution of masses of these three
associations follows a universal mass function, we estimate the number of
members of these associations that remain to be detected. The analysis of rotation and stellar masses using the luminosity X-rays
indicators $L_{x}$ and $L_{x}/L_{b}$ present similar properties, as the dependence on stellar mass and rotation, at least for the younger
associations TWA and BPMG, to those obtained for T Tauri stars in the Orion
Nebula Cluster (1 Myr). A strong desaturation effect appears
at  $\sim$30 Myr, the age of Tuc/HorA, measured essentially by the early G and
late type F stars. This effect seems to be provoked by the minimum
configuration of the stellar convection layers, attained for the first time
for the higher mass stars at  $\sim$30 Myr. The desaturation appears to be independent of rotation at this stage.

\end{abstract}





\section{Introduction}
Studies of the rotational evolution of pre-main sequence (PMS) stars, from the initial stage of T Tauri stars (TTS) up to the arrival to the main sequence (MS), are mainly devoted to angular momentum transfer processes, X-ray emissions and on possible
actions of dynamo mechanisms. Due to the recent use of wide field CCD image applied to dense PMS clusters, a spectacular increase of measured photometric rotation periods with very high accuracy has taken place in the literature. Also, the use of ROSAT and especially CHANDRA X-ray satellites enabled the measurement of hundreds of PMS in the Orion Nebula Cluster (ONC) with an age of 1 - 3 Myr (Feigelson et al. 2003, Flaccomio et al. 20003a, Stassun et al. 2004).

Several challenging problems, such as the existence or not of a bimodal rotation for stellar masses greater than 0.4$M_\odot$ in the ONC (Herbst et al.2001, Rebull et al.2001), have been discussed recently in the literature. The reality of these modes in ONC could be related to the action of a magnetic
braking mechanism between the star and the surrounding accreting disks. In
this sense several studies directed their efforts to study large samples in
the ONC and other similar young clusters. Nevertheless, Stassun et al. (1999) found no
evidence for a dominant magnetic disk locking in ONC. Rebull et al.(2004) proposes the necessity of even larger samples in order to obtain a clear
statistical correlation between rotation periods and disks. Another problem is
the presence of rapid low mass ($\leq$ 0.6 $M_\odot$) rotators in the relative older
cluster of the Pleiades ($\sim$100 Myr) \citep{barnes}.
Stassun et al. (1999) showed that the distribution of the projected rotational
velocities ($vsini$) of Pleiades stars is similar to that of the ONC,
indicating that what exists, in this case at 100 Myr, was already established in
the first 1 Myr (see also Mathieu 2003).

In general, these problems seem to be related to the different lifetimes of
the accreting disks \citep{tinker}. In view of this, it
appears that studies of stellar rotation in the \textit{age - gaps} between 1 and 100
Myr \citep{stassun2003} are necessary and will probably shed more
light on these matters and also on the problem of up to when the
angular momentum is conserved or not \citep{wolff}. By
considering in this work post-T Tauri stars (PTTS) with ages between 8 and 30
Myr, we try to fill these gaps and also to put into evidence some of the
initial speed up processes of stars (Bouvier, Forestini \& Allain 1997; Siess \& Livio 1997) which could begin to occur at $\sim$10 Myr (Rebull et al. 2004). This speed up is possible, because
PTTS, apart some very few exceptions, are devoided of TTS types of disks. These stars are then free to increase their rotations following a normal evolutionary
process.

Concerning the more evolved PTTS with ages 
$\ge$ 5 Myr up to near 40 Myr, measured periods are scarce. One of the reasons for this absence, is the inherent difficulty in measuring these stars which, because of their proximity ($\le$ 100 pc), are distributed in vast areas of the sky. However, some photometric periods have been obtained, for 34 PTTS in Lupus \citep{Wich98}, with a mixture of ages between near 10 and 37 My and also for 9 PTTS belonging to visual binaries Lindroos systems where the primary is a late B star \citep{Nuria02}. In addition, a large mixture of ages is also here present.             

Even if PTTS can be generally defined as low mass stars with ages between 10 and 100 Myr \citep{jen01}, it is not clear what are the
real limits in both extremes. The purpose of this work is to introduce, for the first time as far as we know, a general study of 
rotation of PTTS belonging to coeval nearby  associations of ages between $\sim$10 and $\sim$30 Myr. Contrary to the PTTS mentioned
before, these stars belong to  moving groups having definite ages. We believe that the use of discrete ages systems will introduce a fine-tuning and
a more clear picture of the evolution of PTTS in this interval of ages. To do this, we use the following nearby associations: TWA, BPMG and Tuc/HorA. The ages of these groups have been determined by using a Hertzsprung-Russel (HR) diagram, but in some cases a completely
different and independent method, based on the confinement of past 3D Galactic
orbits, produced similar results confirming these ages. This is the case of
BPMG where the here adopted HR age of 12 Myr (Zuckerman et al. 2001a) is consistent with that obtained kinematically (Ortega et al. 2002, 2004). For TWA, proposed and used ages based on the HR
method range between 5 and 10 Myr (see for instance, Webb et al.1999, Luhman
(2001), Torres et al.2003, Weinberger at al.2004). Here, we adopt for TWA the age
of 8 Myr based on recent results obtained kinematically (de la Reza et al.2004 to be submitted). For Tuc/HorA we adopt the age of 30
Myr, which was independently found to be near 30 Myr by the HR method by
Torres et al. (2000) for the Horologium Association and by Zuckerman et al. (2001b) for the Tucana Association.

Unfortunately, no measurements of periods for individual members of PTTS associations are presently available, but only their projected $(vsini)$ rotation velocities. Because of this, we present here a method to derive representative equatorial velocities $v_{0}(eq)$ from the $vsini$ data in order to estimate mean periods.

The organization of this paper is the following: In Section 2 we derive the
masses and radii and compile a general table containing the main parameters
of all members of the associations that will be used in this work. In Section 3
we describe the method to derive the equatorial rotation velocities from $vsini$
data. In Section 4 this method is applied for the three associations under study. In
Section 5 we discuss some statistical considerations. Section 6 is devoted
to the related X-ray radiation relations and finally the Section 7 contains a summary and our conclusions.

\section{Stellar masses and radii}

Unfortunately, at present, a fundamental calibration of PMS evolution
tracks is lacking. Some of the reasons for this are to be found in the absence of
a large sample of model independent mass determinations. Hillenbrand \& White (2004) have
recently made a careful analysis of masses derived for PMS and MS stars by means 
a complete set of models and their confrontation with available measured 
orbital dynamic data for PMS and early MS stars with masses up to 2.0$M_\odot$.
Consistency appears, in general, to be better for MS stars above 1.2$M_\odot$.
However, for masses below the solar one, PMS models systematically
underestimate the masses by 10\% to 30\%, especially below 0.5$M_\odot$. There are
nevertheless, no dynamical PMS constraints for masses below 0.3$M_\odot$.

Concerning the PMS stars in associations, the situation is worst because in general no mass values have been published. In
these circumstances and as the models give in general similar results we have chosen only one model (Siess, Dufour \& Forestini
2000) to obtain the masses and radii of all the members of the considered
associations at their respective ages. This has been done for masses
comprised between 0.1 and 2.5$M_\odot$ using available observed photometric data of
the individual members. Following the recommendation of Stauffer (2001) we use,
whenever possible, $(V-I_{c})$ colors and when this is not the case, $(B-V)$ colors.

All stellar masses and radii derived this way are listed in Tables 1,2 and 3
together with their observed photometric colors. Some considerations about the
completeness of the members of these associations will be made in Section 5.

\section{The Stellar Rotation}
\subsection{Determination of mean equatorial velocities}

The distribution of projected $vsini$ rotational velocities $\phi(vsini)=\phi(y)$ is related  to the true rotational velocities distribution $f(v)$, considered here to be the equatorial velocities $v_{0}(eq)$, through an integral equation \citep{chan50}. Assuming that the rotation axes
are randomly distributed in space, this relation is

\begin{equation}
\phi(y)=y\int_{y} ^{\infty} \frac{f(v)dv}{v\sqrt{v^2-y^2}}
\end{equation}

It is not an easy task to derive $f(v)$ from $\phi(y)$ by a numerical inversion procedure \citep{gaige93} because $\phi(y)$ has to be differenciated. For this, it is necessary  that the observed distribution $\phi(vsini)$ (normally in histograms) be transformed into a continuous function. This is possible when we have at our disposing a large collection of good quality $vsini$ data. The inversion has been realized for the Pleiades young cluster (age 70-125 Myrs) using $vsini$ data for 235 members stars \citep{All98}. Differently from young clusters, where in general more than one hundred stars can be measured, the PTTS associations contain much less members. This is the case of TWA and BPMG each of which possesses near 30 members with a large proportion of visual binaries in the case of TWA.
Adding all probable and possible members of Hor and Tuc A (where much less binaries are present) we obtain a total of about 50 objects.

Due to the relative low number of PTTS in associations and also considering the uncertainties in $vsini$ values between 2 and 5 $km$ $s^{-1}$ (in the best case) for low and high rotators respectively, a different strategy must be followed. We follow here the suggestion of Chandrasekhar and Munch (1950) for the case of few data which consists of assuming a parametric form of $f(v)$ instead of inverting the integral equation (1). We can then compute the integral in (1) and obtain the corresponding $\phi(vsini)$ distribution which will be compared to the observational distribution of $vsini$ for each association.

Different functions, from a simple rectangular to a more elaborated parabolic representation were tested by Brown (1950) and give, so as the gaussian used by Chandrasekhar and Munch, fairly similar results. We then use this rectangular function as proposed by Brown by taking $f(v)dv$, the probability of ocurrence of $v$ between $v$ and $v+dv$, to be equal to $dv/2a$ for $v>v_0-a$ and $v<v_0+a$ and equal to zero otherwise. The rectangle centered at the probable velocity 
$v_0$  has a height of $1/2a$ and a total width of $2a$. To compare with observations, a cumulative distribution function of the projected velocities is defined as $\Phi(y)=\int\phi(y)dy$. The integration is realized over finite intervals of $y$ corresponding to the grouping of the observations. The fraction of stars with $(vsini)_l<y<(vsini)_{l+1}$ is given by $\Phi(vsini)_{l+1}-\Phi(vsini)_l$. Multiplying this difference by the total number of stars in the association we obtain the total number of stars in that interval of $vsini$ values. The resulting cumulative function $\Phi(y)$ is a monotonically increasing function from zero to one as $y$ increases. To calculate $\Phi(y)$ we used the relation (6) of Brown (1950).
 The derivative of $\Phi(y)$ using one hundred successive intervals of $y$ will finally be compared with the observed $\phi(vsini)$ distributions.

\section{Observed rotation data}

The start point of our analysis will be stellar masses, more properly we will choose \textit{a priori} a specific mass value in the interval 0.1-2.6$M_\odot$ separating low mass (LM) and high mass (HM) stars. This mass divisor is chosen by considering the $vsini$ distribution as presented in Figures \ref{fig4} and \ref{fig5}.
 
An examination of the points in the ($vsini$, mass) plane of these two figures permits to distinguish that a different pattern of $vsini$ values appears at $\sim$1.5$M_\odot$. Because of this, we call LM stars those with masses smaller than 1.5$M_\odot$ and HM stars, those with masses larger or equal than 1.5$M_\odot$. The corresponding low mode (LRM) contains, in general, low $vsini$ values (note however, that Tuc/HorA contain some LM stars with high $vsini$ values). The high rotation mode (HRM) contains higher $vsini$ values as compared to the low ones. As will be seen in section 5, this choice of mass separation finds a statistical justification. Anyway, we expect that the physical importance resulting from the above separation of stellar masses will emerge in this work.

\subsection{The Spin up and the specific angular momentum behavior}

\begin{center}
\begin{figure}
\begin{center}

\includegraphics[angle=0,width=17cm]{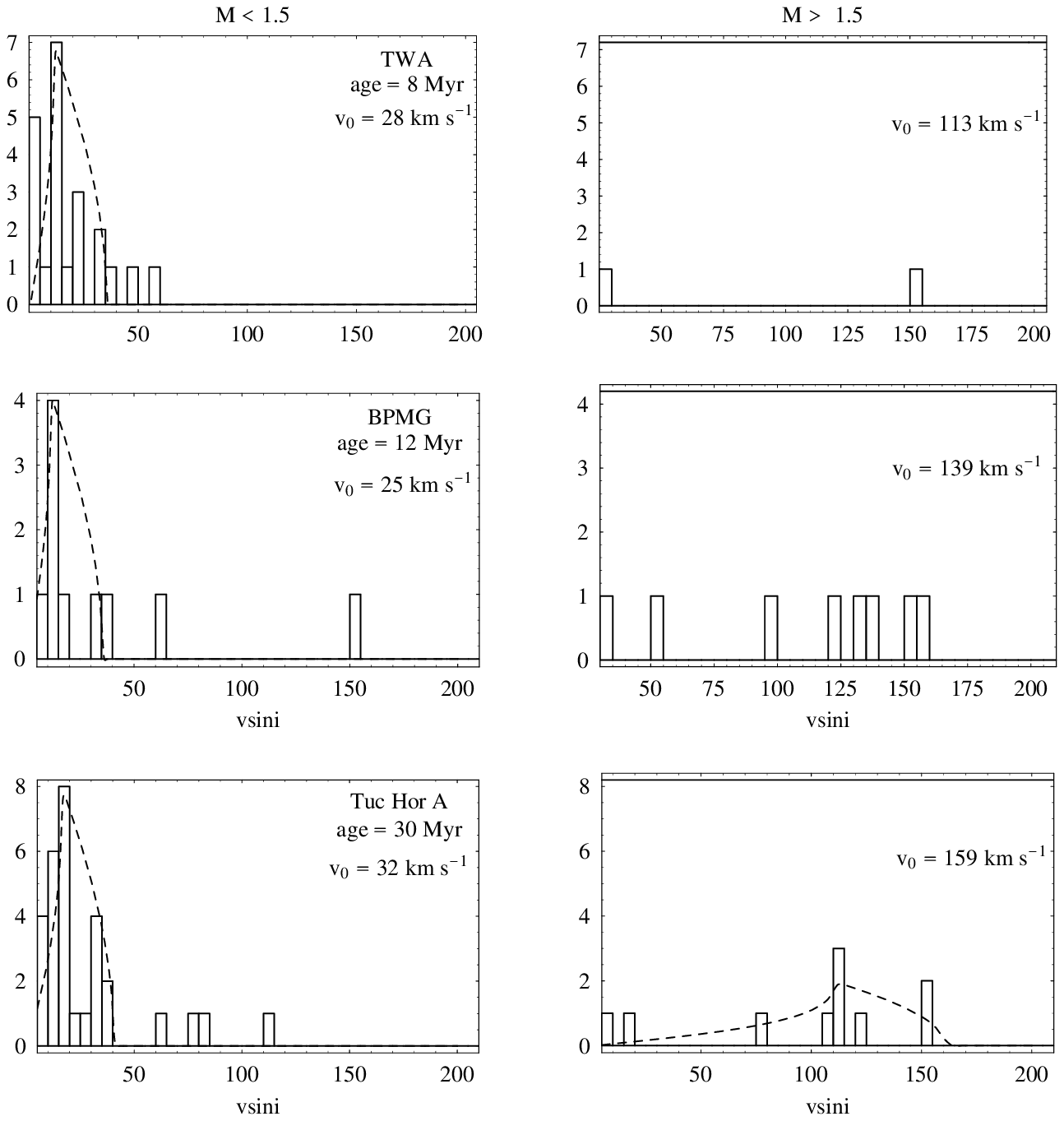}
\caption{$vsini$ distributions in $km$ $s^{-1}$ for the three associations studied in this work. Left column correspond to masses less than 1.5$M_{\odot}$ and right column to masses greater or equal than 1.5$M_{\odot}$. Mean equatorial rotation velocities $v_{0}(eq)$ for LM and HM rotation modes corresponding to the fitting procedure (see text) and the ages for each association are indicated. We have taken $a=5$ $km$ $s^{-1}$ except for the high mode of Tuc/HorA in which  $a=10$ $km$ $s^{-1}$}
\label{fig1} 
\end{center}
\end{figure}
\end{center}

In Figure \ref{fig1} we present six histograms containing the distributions of the observed $vsini$ values for the LRM and HRM and for the three considered associations. The dashed line curves in some of these histograms represent the best fit of the frequency distributions. For the LRM they correspond to mean equatorial velocities $v_{0}(eq)$ of 28$\pm$5, 25$\pm$5
and 32$\pm$5 $km s^{-1}$ respectively for TWA, BPMG and Tuc/HorA. For the HRM we were able to adjust a curve only for the case of Tuc/HorA where a small peak appears to emerge. Due to the nature of the HRM $vsini$ distribution of the HRM of TWA (where only two HM stars have been detected until now) and BPMG, our fitting procedure does not apply. Nevertheless, quite good representative $v_{0}(eq)$ values can be obtained by multiplying average $vsini$ value by the factor $4/\pi$ \citep{chan50}.

In these conditions, the resulting HRM $v_{0}(eq)$ values for TWA, BPMG and Tuc/HorA are respectively equal to 113, 139 and 159$\pm$5 $km s^{-1}$. All  $v_{0}(eq)$ values are indicated in Figure \ref{fig1} and in Table 4. Considering the successive ages of these associations, a mean spin up appears to be present for the HRM, whereas the $v_{0}(eq)$ in LRM are not so different. This behavior agrees with the known slow stellar evolution of LM stars contrary to HM stars, where the evolution is faster. Future measured periods of individual star members of these associations, will probably allow to confirm these results. In any case, we found a continous spin up behavior up to 30 Myr which is near the 40 Myr maximum spin up found in PMS models (Siess \& Livio 1997). If this is the case, HM stars could be used as gyrochronometers following the spin pattern of different associations. A new tool would then be at hand to estimate the age of an association by examining their high mass content.

By estimating a mean radius $<R>$ and using the $v_{0}(eq)$ mentioned above for these associations we can determine mean rotation periods as we will see later,
but for the moment we can estimate the behavior of the specific mean angular momentum $J/M=k_{tot}^{2} \times v_{0}(eq) \times <R>$, where $k_{tot}$ is the total radius of gyration. Depending on the mass and on the age, the values of $k_{tot}^{2}$ consider the changes of the stellar moment of inertia, specially the receding of the convective zone leaving more centrally condensed stars. Here, we have used the $k$ values for convection and radiation furnished by the PMS models of Siess et al.(2000). The ensemble of $k^{2}$ values and the resulting $J/M$ values in  $cm^{2} s^{-1}$ for the representative $v_{0}(eq)$ and $<M>$ are presented in Table 4. Taking into account the adverse effects of an unequal mass distributions of these associations, the fact that only two HM stars are present in TWA, and the dominant large radii estimated in BPMG (see Table 2), we found that a mean decrease of factor two is present for low mass stars between $\sim$10 and 30 Myr. Some undetected mechanisms is active in the low mass stars dissipating angular momentum and preventing these stars from having sufficiently larger surface velocities in order to conserve their $J/M$. On the other hand, the high mass stars develop a sufficient spin up that practically mantains a conserved $J/M$ in this same time interval.  Also, with the exception of the LRM comparison between TWA and BPMG, the other three possible comparisons between the modes and associations of different ages show that $<R>$ decreases as $\propto t^{-1/5}$.

\section{A statistical approach}

In Section 4 by means of the observed $vsini$ distribution we found, for the three associations,  mean representative equatorial rotation velocities for two different types of stellar masses. Also we found that a clear spin up is present, at least for the considered high mass stars between 8 and 30 Myr (see Table \ref{tbl-med}). In this section we explore the completeness of our data members and perform a statistical treatment of the mass and $vsini$ data in order to examine the consistency with the observed spin up process.

\subsection{Initial Mass Functions}

The number of members of the associations collected in Tables \ref{tbl-TWA}, \ref{tbl-BPMG} and \ref{tbl-TucHorA} is not intended to be complete. We have selected from the literature all probable and possible members, although is quite possible that future investigations could eliminate some of these members. On the other hand, several members depending on their masses, remain to be detected. In order to estimate the number of stars that probably could be discovered, we assume that the mass spectrum follows an universal mass distribution. We do so, even considering that the formation of these groups could be triggered by a mechanism, different from those of young clusters related mainly to gravitational processes. In fact, as shown in Ortega et al (2002, 2004), BPMG was probably formed $\sim$12 Myr ago by the impact pressure of a supernova front shock originated in the OB association Sco-Cen. This appears also to be the case, for TWA $\sim$8 Myr ago (de la Reza et al. 2004, to be submitted).

In this discussion we consider only stars with masses between 0.1 and 2.6$M_\odot$ that we obtain in these associations. We do not discuss, or take into account, the few brown dwarfs stars with masses less than 0.1$M_\odot$ that have been detected among these groups. This is the case of TWA 5B, GSC 8047-0232B and HR7329B (see for instance Neuh$\ddot{a}$user \& Guenther 2004). Also, visual binaries (TWA is full of them!) are considered as simple stars for mass distribution purposes.

As a convenient \textit{universal} initial mass function (IMF) we choose the one proposed by Scalo (1998) (see below), but before adjusting it to the data we have to take into consideration that the observed mass distributions in the associations are slightly different. While, TWA appears to be well fitteded in the range of very low mass stars (0.1-0.25)$M_\odot$, this is not the case for BPMG and Tuc/HorA. On the contrary, higher masses appears to be better represented in BPMG, Tuc/HorA than in TWA. This last group has only two stars in our defined high mass regime.

\begin{center}
\begin{figure}
\begin{center}

\includegraphics[angle=0,width=12cm]{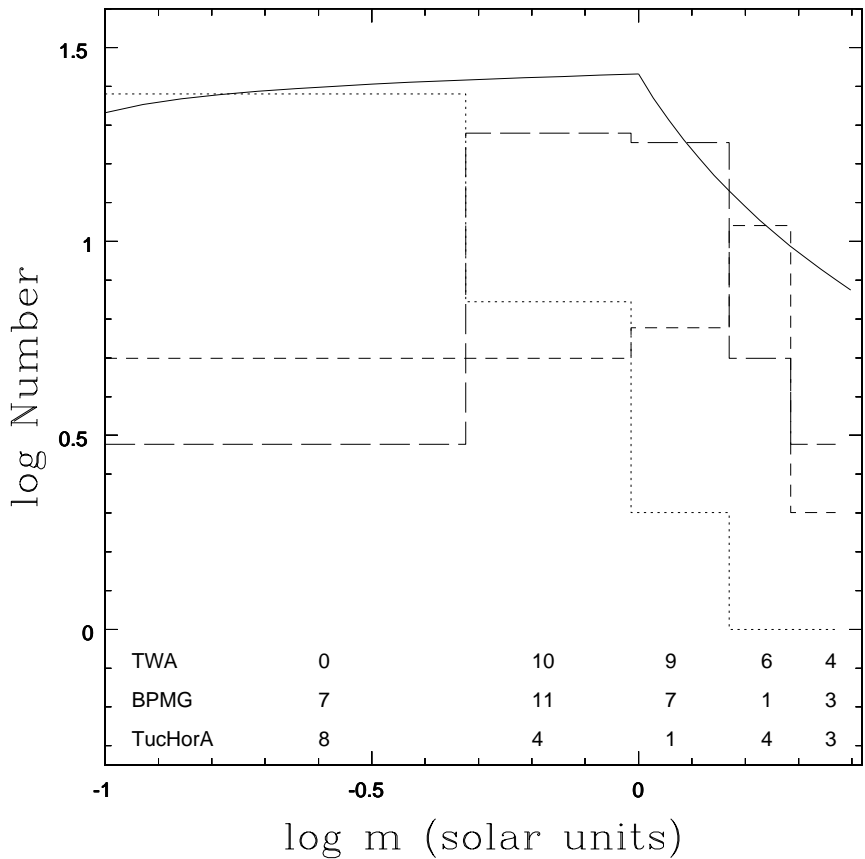}
\caption{Comparison of mass distributions with the IMF. The dotted histogram correspond to TWA, long dashed histogram to BPMG and the short dashed for Tuc/HorA. Our two-segment power law are represented by a continuous line. The number of predicted stars for each association and mass intervals obtained from the differences between the IMF and observed distributions is indicated.} 
\label{fig2}
\end{center}
\end{figure}
\end{center}

Here, we assume that TWA is complete in the very low mass range. We can then fit the height of Scalo distribution with this group. The adopted Scalo (1998) distribution is the following:

\begin{eqnarray}
\xi(m) & = k\left\{
\begin{array}{cc}
D_{0}m^{D_{1}} & if  \;\;\; 0.1 \le m/M_{\odot} \le 1.0\\
D_{0}m^{D_{2}} & if  \;\;\; 1.0 \le m/M_{\odot} \le 2.6\\
\end{array} 
\right\}
\end{eqnarray}

where $k$ is an adjusting constant, $D_{1}$ and $D_{2}$ correspond to the slope of logarithmic IMF
at each mass interval and $D_{0}=32$ is a constant involving the age of the
Galactic disc \citep{miller}. Analytic fits to the IMF using only two parts of the three-segment power law given by Scalo (1998) give: $D_{1}=-0.2\pm0.5$ and $D_{2}=-1.7\pm0.5$, corresponding to the mass interval 0.1-10$M_\odot$. We have obtained the best fit to the observations by adopting $D_{1}=0.1$ and $D_{2}=-1.4$. In Figure \ref{fig2} we show the result of this fit for the three associations and for which we have adopted a general bin of 0.5$M_\odot$.

The number of stars that remain to be discovered for each association, if they follows the IMF, can be estimated by the difference of the number of stars between the IMF and the observed distributions. These are shown in Figure \ref{fig2}. A first tentative to estimate the number of high mass stars of TWA by a IMF approach was made by \citep{web01}.

\subsection{$vsini$ distribution with age}

\begin{center}
\begin{figure}
\begin{center}
\includegraphics[angle=0,width=10cm]{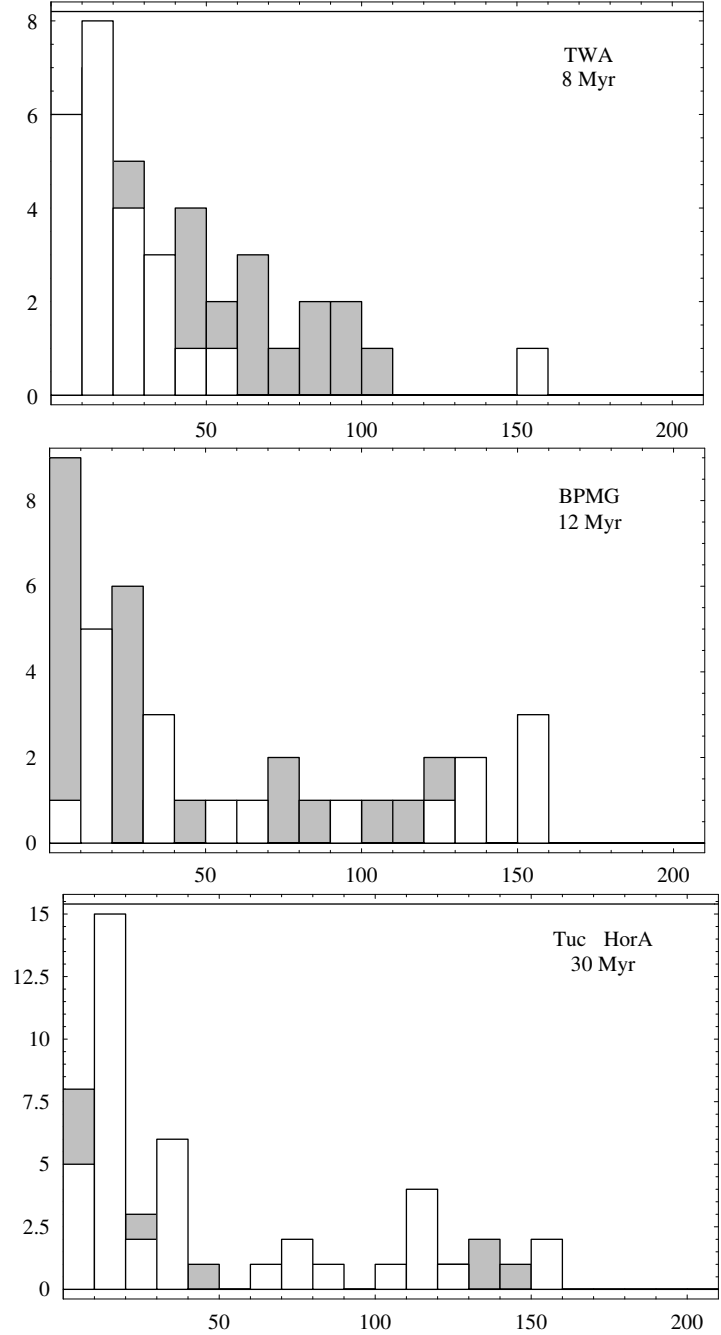}
\caption{$vsini$ distributions in $km$ $s^{-1}$ of the LM and HM rotating modes. White and grey bars represent respectively observed and simulated values (see text)} 
\label{fig3}
\end{center}
\end{figure}
\end{center}

Having of a number of observed stars with their respective masses and $vsini$ values and using also the number of the remaining stars to be detected, we can try to complete the observed histograms presented in Figure \ref{fig1}. For this we must simulate $vsini$ values for these remaining stars. This can be done by multiplying the mean $v_{0}(eq)$ values obtained from observational data in Section 3 by random $sini$ values (note that this assumption of random distribution of the inclination axes has been made to obtain the $v_{0}(eq)$ values). Several tests of random $sini$ between 0 and 1 produced similar results because they are conditioned by the used  $v_{0}(eq)$ values. A typical result is shown in Figure \ref{fig3}. We notice that the distribution of observed and simulated values of $vsini$ is compatible with a spin up process in the interval of ages. We should note that we cannot use the evolution of this distribution to extrapolate to younger and older ages to obtain a ressemblance of the $vsini$ distribution as is the case for comparison between the distribution of $vsini$ in ONC and the Pleiades done by (Stasssun et al.1999, see also Mathieu et al.2003). This is because these associations do not appear to contain very high rotators with masses below 0.6$M_{\odot}$ as in the case of the Pleiades.  Nevertheless, some observed low mass with very high $vsini$ values exist in Tuc/HorA for masses between 0.7 and 1.0$M_{\odot}$ (see Figure \ref{fig5}). Also, our determination of mean equatorial velocities is not capable to properly consider these few stars.

\subsection{A two dimensional K-S analysis}

In order to statistically quantify the observed distribution of $vsini$ for low and high mass stars between two associations with different ages we use a Kolmogorov-Smirnov analysis (K-S). While the standard one dimension K-S test does not appear to produce significative measurable differences, with the exception of the obvious case of HRM between TWA and BPMG, a two dimensional K-S test appears to be more interesting to visualize the spin up process.

Differently from the traditional one dimension K-S test, where a simple value can represent the maximum difference between two cumulative distributions, the two dimensional K-S test use a relative different approach. This is due to the fact that a cumulative two dimensional distribution is not defined. Following Press et al. (1994) the discriminator must then be viewed in one quadrant of the four containing all the points in a two dimensional distribution.

\begin{center}
\begin{figure}
\begin{center}

\includegraphics[angle=0,width=14cm]{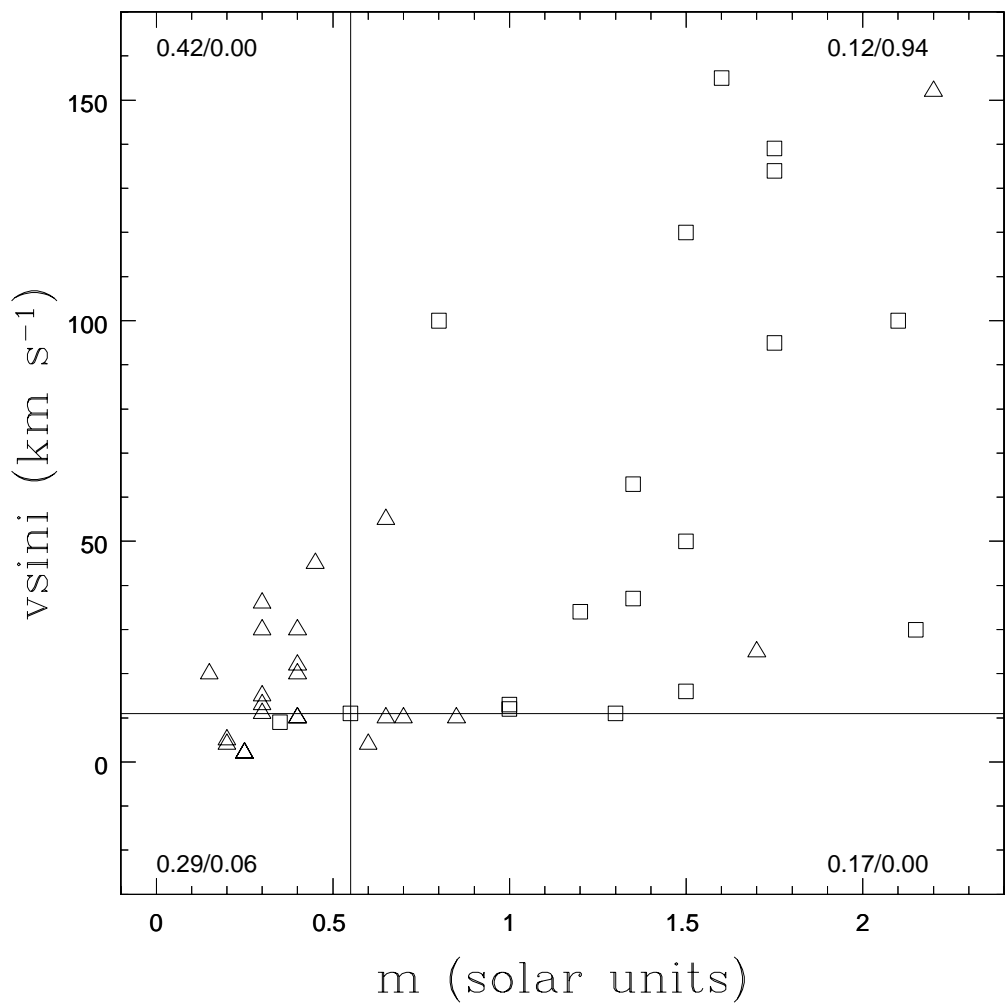}
\caption{Two dimension K-S test applied to the mass-$vsini$ distributions corresponding to the TWA and BPMG associations. The algorithm of Press et al.(1994) calculates the origin point whose
quadrants maximizes the difference between fraction of triangles (TWA) and fraction of squares (BPMG). The maximum occurs in the upper-right quadrant. At this quadrant the smallest probability of null hypothesis is found.} 
\label{fig4}
\end{center}
\end{figure}
\end{center}

\begin{center}
\begin{figure}
\begin{center}

\includegraphics[angle=0,width=14cm]{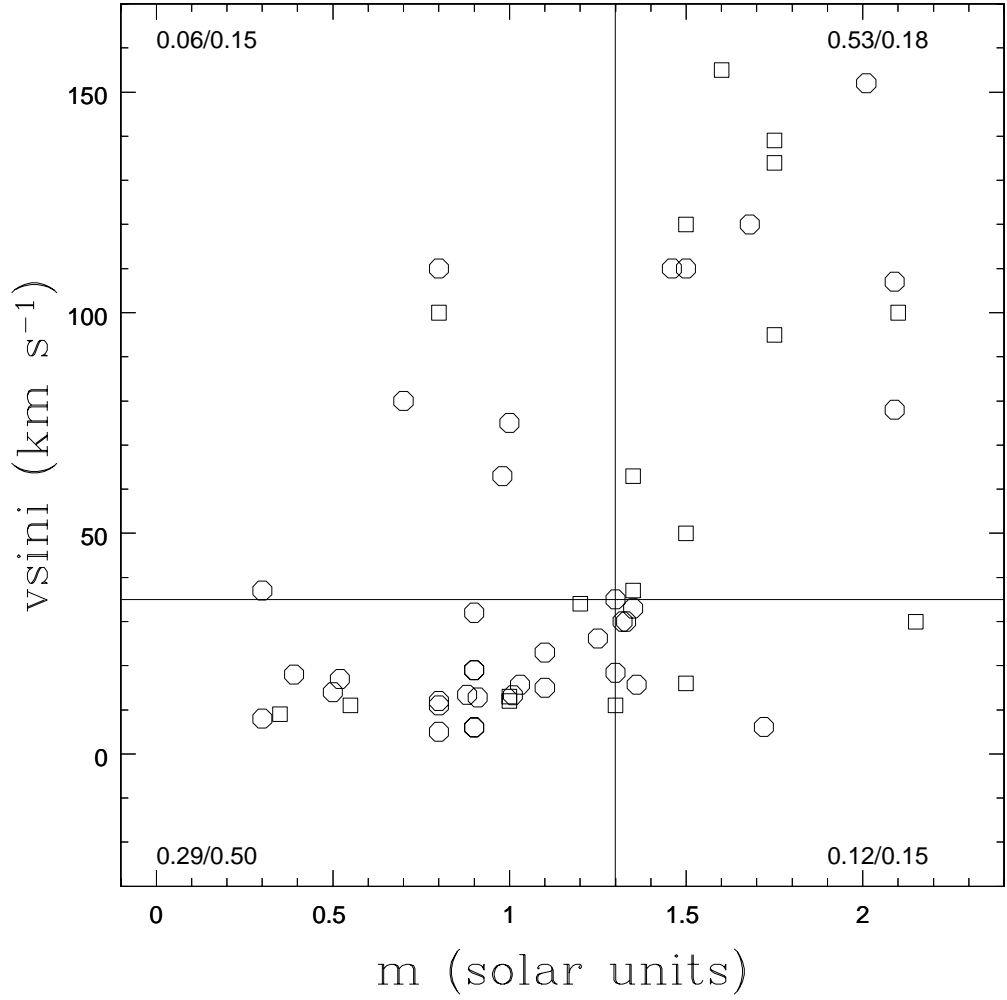}
\caption{K-S test applied to BPMG (squares) and Tuc/HorA (polygons) associations. The same as in Figure \ref{fig4}, the maximum differences are found in the upper right quadrant.} 
\label{fig5}
\end{center}
\end{figure}
\end{center}

The two dimensions chosen here are the stellar mass and the observed $vsini$ values as shown in Figures \ref{fig4} and \ref{fig5} for the TWA-BPMG and BPMG-Tuc/HorA comparison respectively. For the two dimensional K-S test we apply the mathematical recipe presented in Press et al. (1994).

When the populations of two associations of different ages are compared in the same plane, the program evaluate, for each population an origin point of the four quadrants. In principle one of the four quadrants will show the largest differences between the distribution of points. These differences can then be quantified in the following manner: in each quadrant the number of points of each population is counted and divided by the total number of points of each population (without considering the origin point). For each quadrant these two predictions are subtracted. The quadrant that presents the largest difference is the one indicating the largest differences in the points distribution.

In Figure \ref{fig4} for instance we present the results of the comparison of TWA (triangles) and BPMG (squares). Each one producing a system of four quadrants. We represented here only the case centered on BPMG stars (that based on TWA stars, and not showed here, produces similar results). In each quadrant the difference of their corresponding ratios one with respect to the other is shown. The upper value for the TWA-BPMG comparison and the lower one for the BPMG-TWA comparison.

In both cases, we see that the upper right quadrant corresponding to larger masses and higher $vsini$ values is that presents the largest different distribution of values. In Figure \ref{fig5} we show the same type of results for the comparisons between BPMG-Tuc/HorA also indicating the same upper right quadrant for the largest differences.

The axis center is placed at 0.7$M_{\odot}$ when using the TWA distribution because this group has the largest distributions of low masses. Whereas, due to the relatively more massive distributions in BPMG and Tuc/HorA their center point is at 1.3 $M_\odot$. It is interesting to note that this center is similar to our 1.4 $M_\odot$ value used to discriminate LM and HM stars. This supports, in a certain way, our \textit{a priori} chosen mass divisor value.

We performed similar calculations (not shown here) using the observed and simulated $vsini$ values as computed before. The results for the BPMG-Tuc/HorA population comparisons are similar to those using only observed values. The results of the observed plus simulated comparison between TWA and BPMG are however less clear. The upper right quadrant becoming less important and being almost similar to the upper left quadrant. This is due to the largest increase of simulated stars with masses larger than 0.25$M_\odot$ for TWA.

In any case, the two dimensional K-S test for the case of pure observed values is compatible with the presence of a spin-up process dominated by HM stars in these associations.

\section{X-rays versus rotation}

One of the main purposes of studies of X-ray radiation in PMS
stars is the search for correlations with stellar rotation in order to
shed some light on the question about what kind of dynamo is into action (see Feigelson et al. 2003 for a discussion on different dynamo mechanisms). Before
presenting our results concerning the X-rays properties in PTTS for ages
between 8 and 30 Myr we present here the results obtained for ONC ($\sim$1 Myr)
in order to see for possible connections among these ages.

\begin{center}
\begin{figure}
\begin{center}

\begin{tabular}{c}

\includegraphics[angle=0,width=10cm]{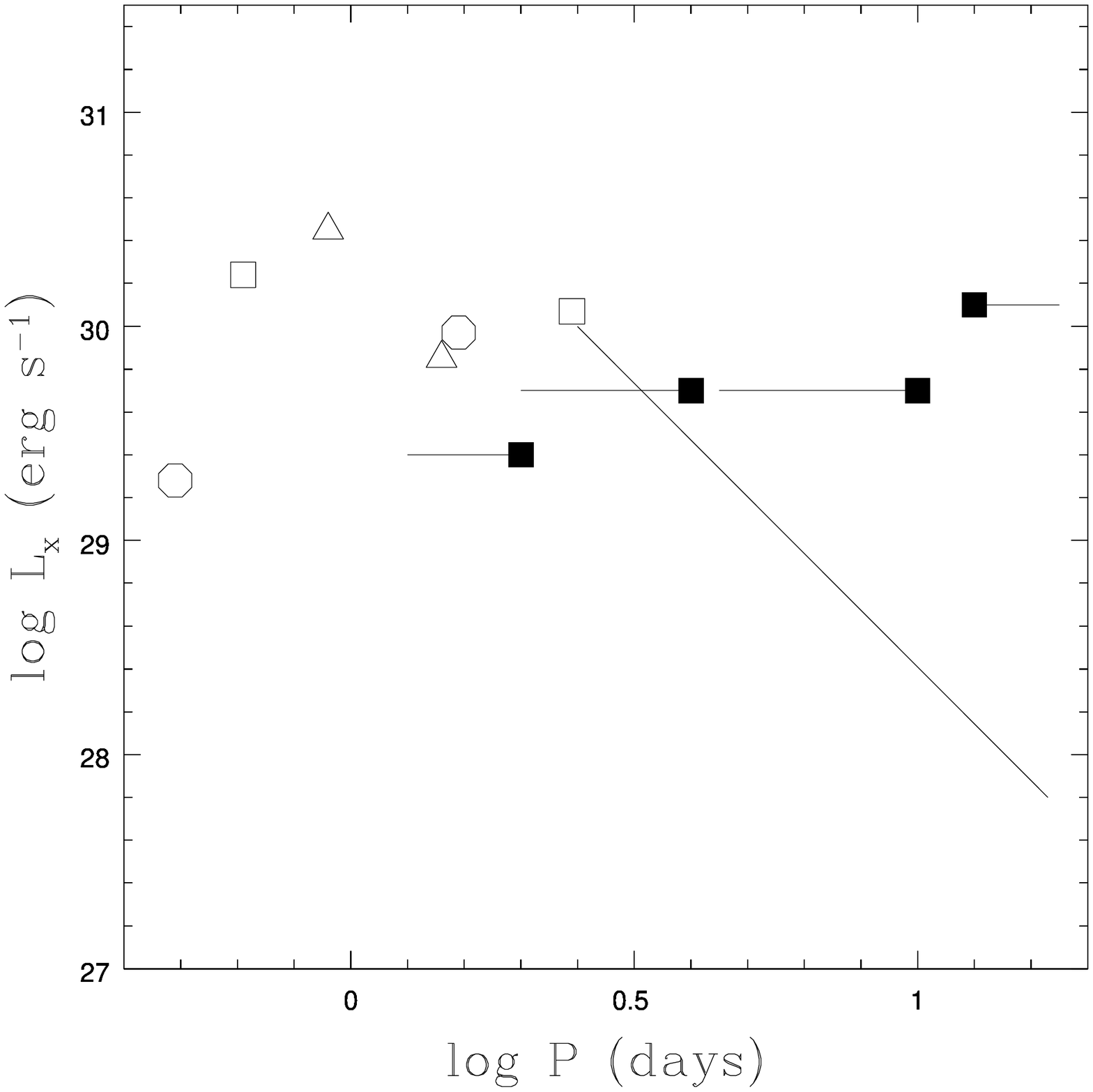} \\
\includegraphics[angle=0,width=10cm]{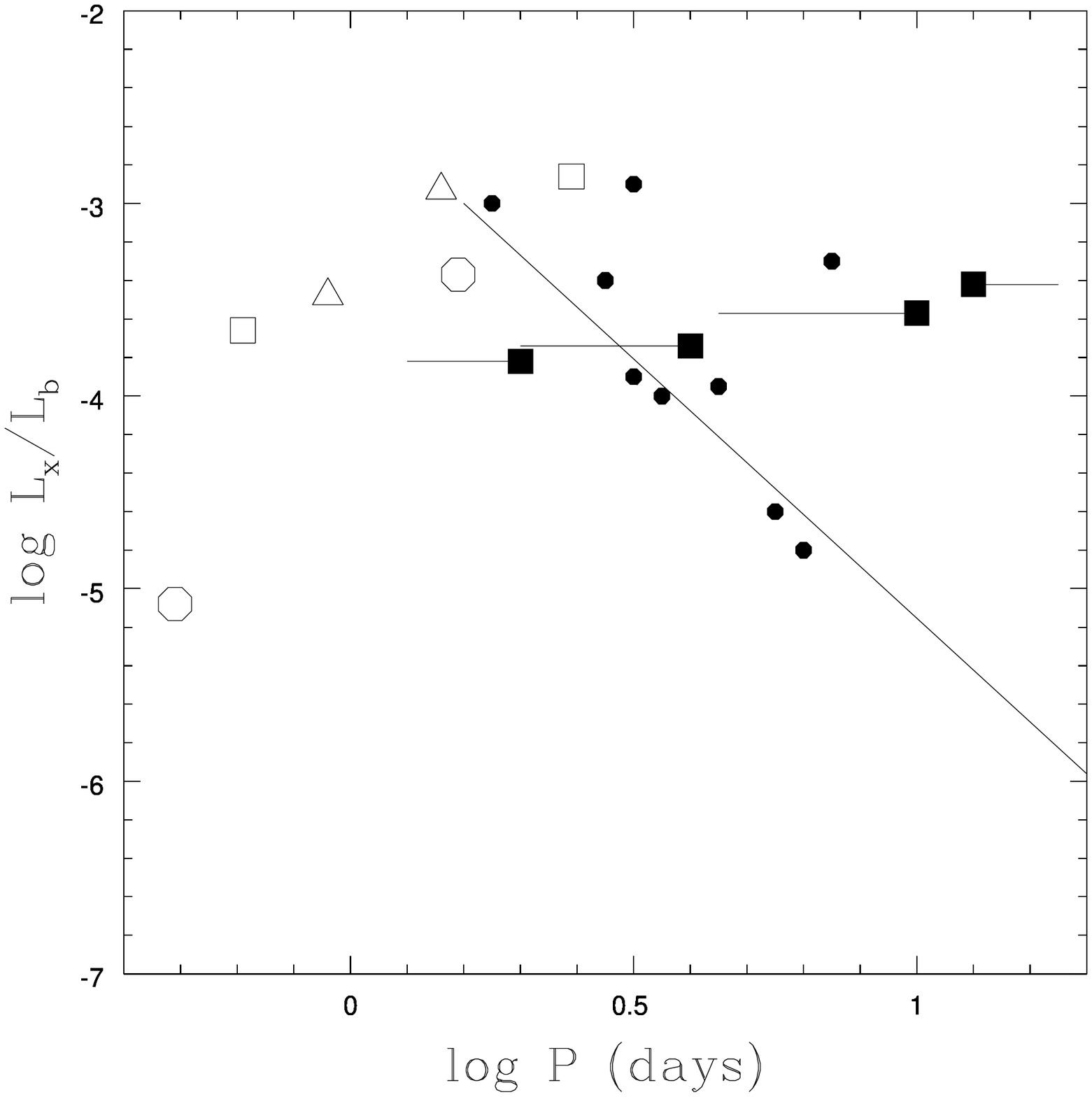}\\
\end{tabular}
\caption{Schematic time evolution of mean $L_{x}$ and $logL_{x}/L_{b}$ for both LM and HM rotating modes. Open symbols represent TWA (triangles), BPMG (squares) and Tuc/HorA (polygones). Filled squares corresponds to ONC data taken from Feigelson et al.(2003). In the bottom panel the filled circles are taken from Hu\'elamo (2002) and represent 9 PTTS belonging to Lindroos systems.} 
\label{fig6}
\end{center}
\end{figure}
\end{center}

Three main recent studies in ONC, based on hundreds of X-rays measurements
obtained with CHANDRA, deserve particular consideration.  Feigelson et al.2003 and Flaccomio et al.2003a found in ONC that stars with measured periods are near
the saturated regime at $log L_{x}/L_{b}$ =$\sim$-3. (-3 is the canonical saturation
value of the linear MS relation shown in Figure \ref{fig6} and $L_{b}$ is the bolometric
luminosity). However, the mean $log L_{x}/L_{b}$ values of Feigelson et al. (shown in Figure \ref{fig6}) present values smaller than -3, suggesting more a super-saturated regime.
Concerning the relations of these fluxes with rotation and differently from
Flaccomio et al. which do not find any apparent rotation/X-ray
relationship, Feigelson et al. does find a slight mean relation with rotation
periods, in an inverse way as that of the MS linear relation. As can be seen
in Figure \ref{fig6}, the Feigelson et al. data appears to show lower $log L_{x}/L_{b}$ values for high
rotators. In a recent work Stassun et al.(2004) re-analyses all the
ONC data, this time including also those stars with no measurable rotation
periods. The main recent results of Stassun et al., which can be compared with those of Feigelson et al. and Flaccomio et al. are the following: 1) most stars with measured
rotation periods (more X-ray luminous in general) are placed in the super-
saturated regime 2) the relation with rotation is confirmed in the sense that
high rotations indicate lower $L_{x}/L_{b}$ values 3) as also found by Feigelson et al.
and Flaccomio et al., $L_{x}$ increases with the stellar mass 4) by finding that
stars with accreting signatures present lower X-ray luminosities, probably
resulting from absorption by the circumstellar disks, they confirm the
Flaccomio et al. results that the non accreting stars have much larger X-ray
luminosities than the accreting ones. 5) an emerging linear MS relation can be
present in stars with no observable rotation periods.

There is then a general indication that central stars, and not the disks, are
the main source of X-rays. We have chosen also
to explore the luminosity $L_{x}$ and the ratio $L_{x}/L_{b}$, as the main X-rays
indicators for the PTTS associations. All these values are taken from ROSAT
measurements when
available. We prefer to use, as far as possible, published $L_{x}$ and $L_{x}/L_{b}$
values. When this is not the case, we calculate these values using published
stellar distances of all the members of the associations. These distances were
furnished  by Hipparcos or by appropiate kinematical and age considerations
used to establish these associations (see Tables 1-3). For the calculations of the fluxes we used classical procedures with $F_{x} = C(8.31 + 5.30 \times HR1)\times 10^{-12}$, where $C$ is the ROSAT counting rate and $HR1$ the hardness ratio \citep{jensen} and with $F_{b}=2.48 \times 10^{-5} \times 2.512\times^{-(V+BC)}$, where $V$ is the observed visual
magnitudes (see Tables 1-3) and $BC$, the bolometric correction estimated using
\citep{kenyon} tables. Due to ROSAT resolution, when visual binaries are very close (excepting when the companion is an A type star) we divided $F_{x}$ by 2. All these $L_{x}$ and $L_{x}/L_{b}$ values are shown in Tables 1-3.

In order to ascertain the general behavior of the X-ray radiation with rotation,
we estimate representative rotation periods using the $v_{0}(eq)$ and mean radii
discussed in Sections 2 and 4. The values of these periods are shown in Table
4. Also in this table are included the mean $L_{x}$ and $L_{x}/L_{b}$ values for the LRM
and HRM for the three associations. A graphic representation of these values
is in Figure \ref{fig6}. This figure contains, for comparison
purposes as mentioned before, a schematical representation of the X-ray
behavior of ONC (Feigelson
et al. 2003). They also contain the linear trend corresponding to the ($\alpha$-$\omega$)
dynamo valid for low rotating MS stars. We also consider in Figure \ref{fig6}, the
tendency found for 9 PTTS belonging to Lindroos systems by Hu\'elamo (2002). Even
if these stars, apparently mimic the
($\alpha$-$\omega$) dynamo, this trend is not due to this effect, but to an ageing
effect. In fact, the upper stars are very young stars whereas the lower ones
are older than 50 Myr.

\begin{center}
\begin{figure}
\begin{center}
\begin{tabular}{c}

\includegraphics[angle=0,width=10cm]{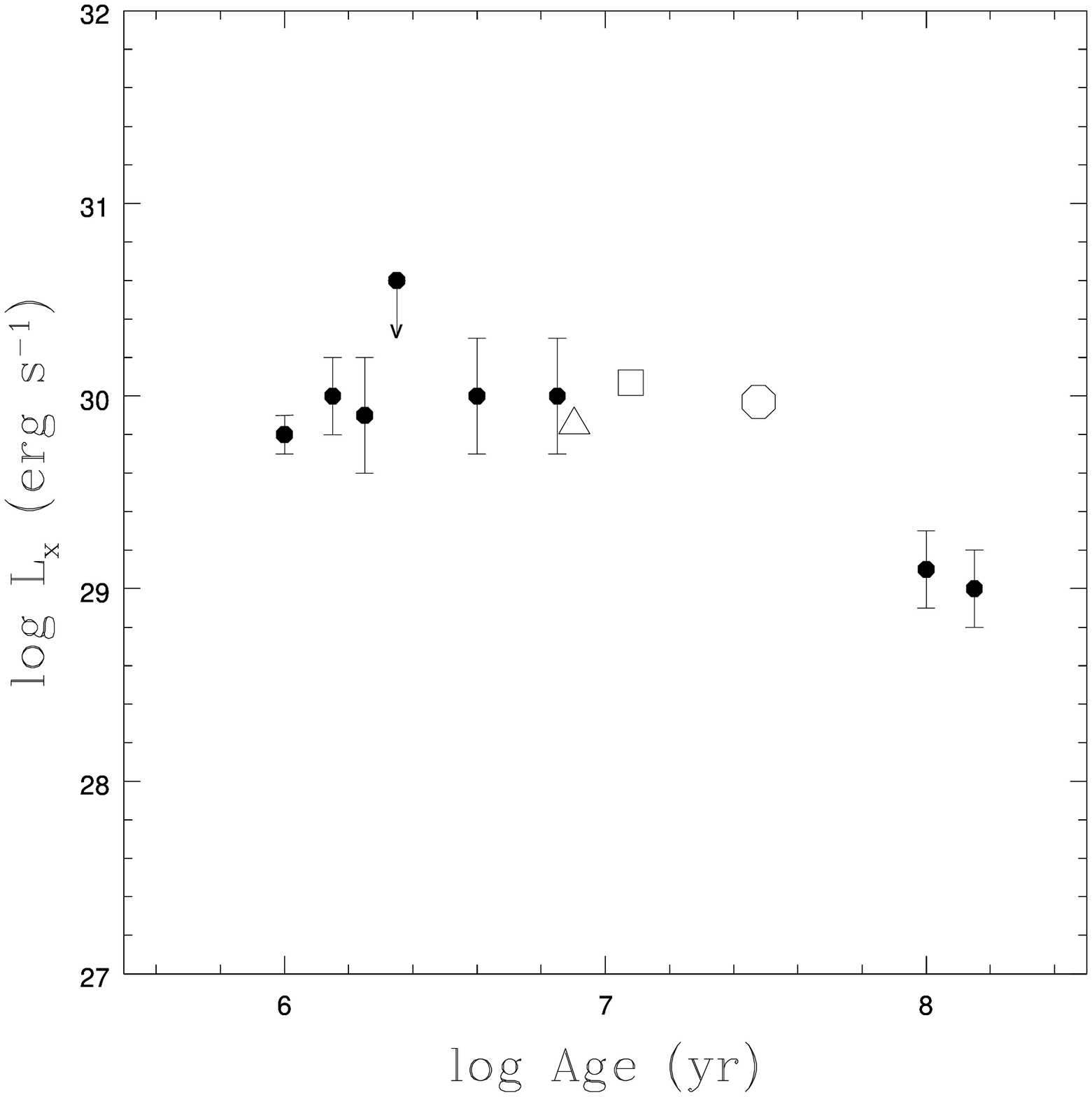} \\
\includegraphics[angle=0,width=10cm]{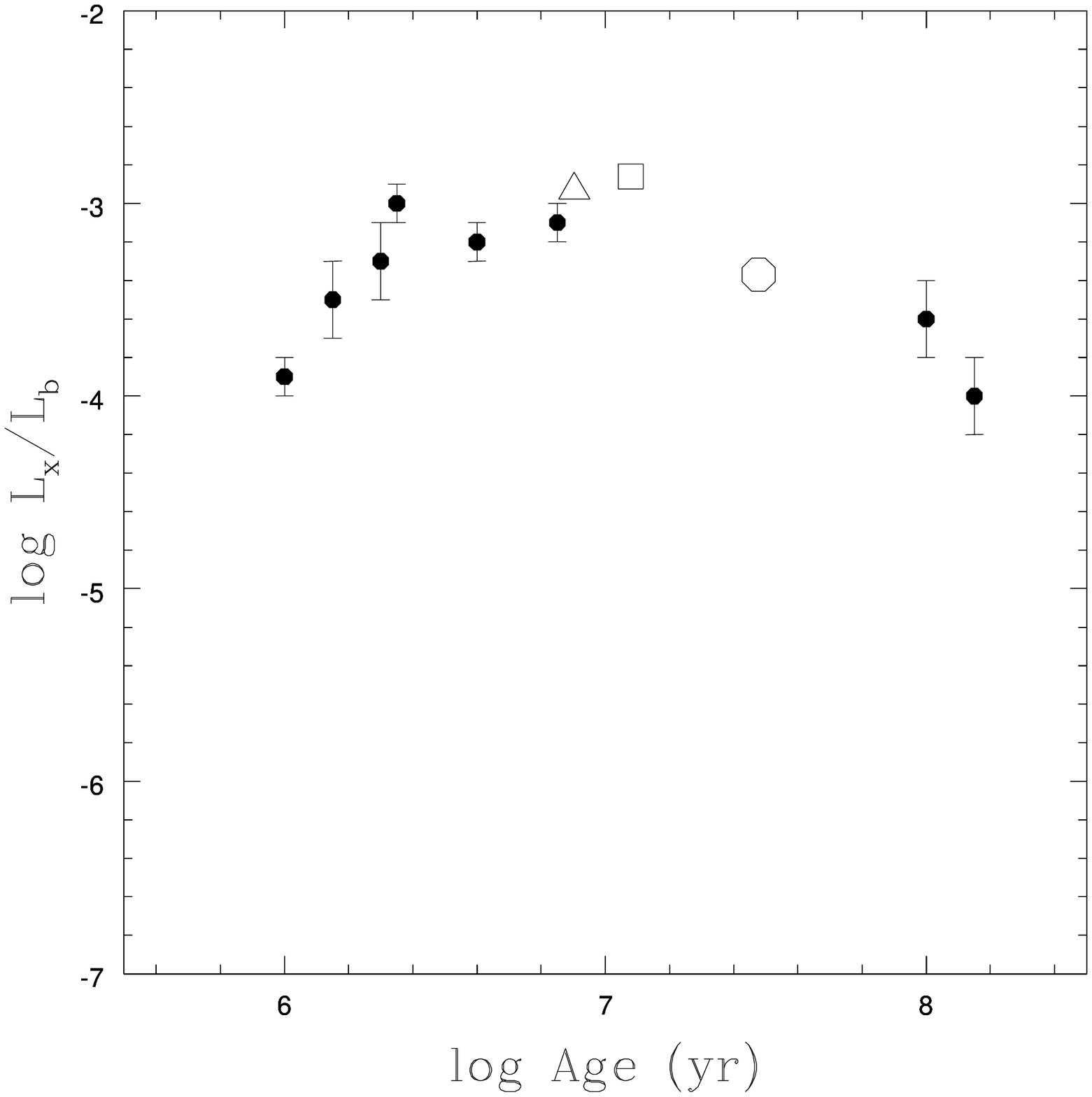}\\
\end{tabular}
\caption{Mean values of the X-rays indicators as function of age for the lower rotating mode. Open symbols represent TWA (triangles), BPMG (squares) and Tuc/HorA (polygones). Filled circles taken from Flaccomio et al.(2003b) represent T Tauri stars (left side) and early MS stars (right side).} 
\label{fig7}
\end{center}
\end{figure}
\end{center}

\begin{center}
\begin{figure}
\begin{center}
\begin{tabular}{c}

\includegraphics[angle=0,width=10cm]{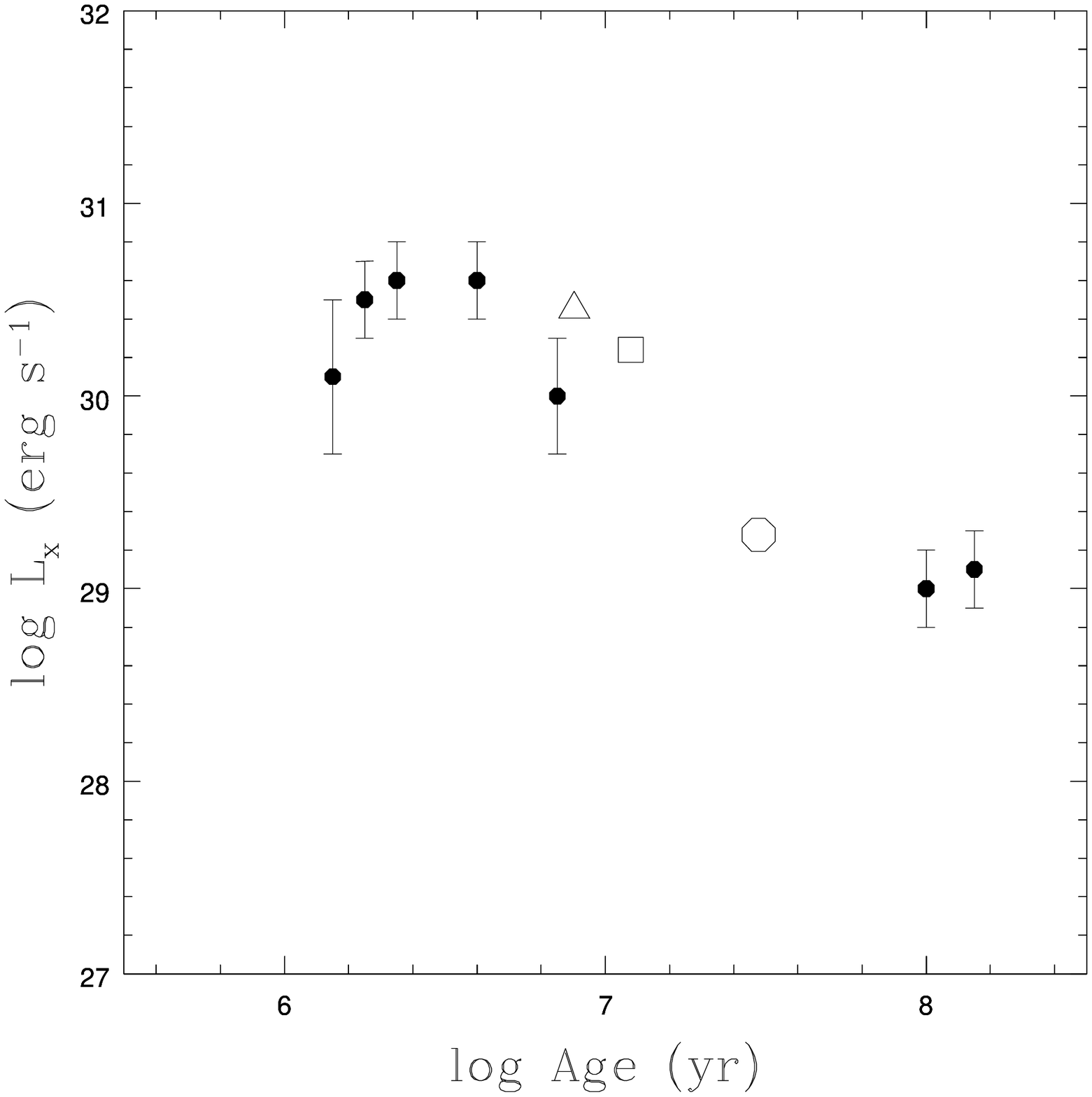} \\
\includegraphics[angle=0,width=10cm]{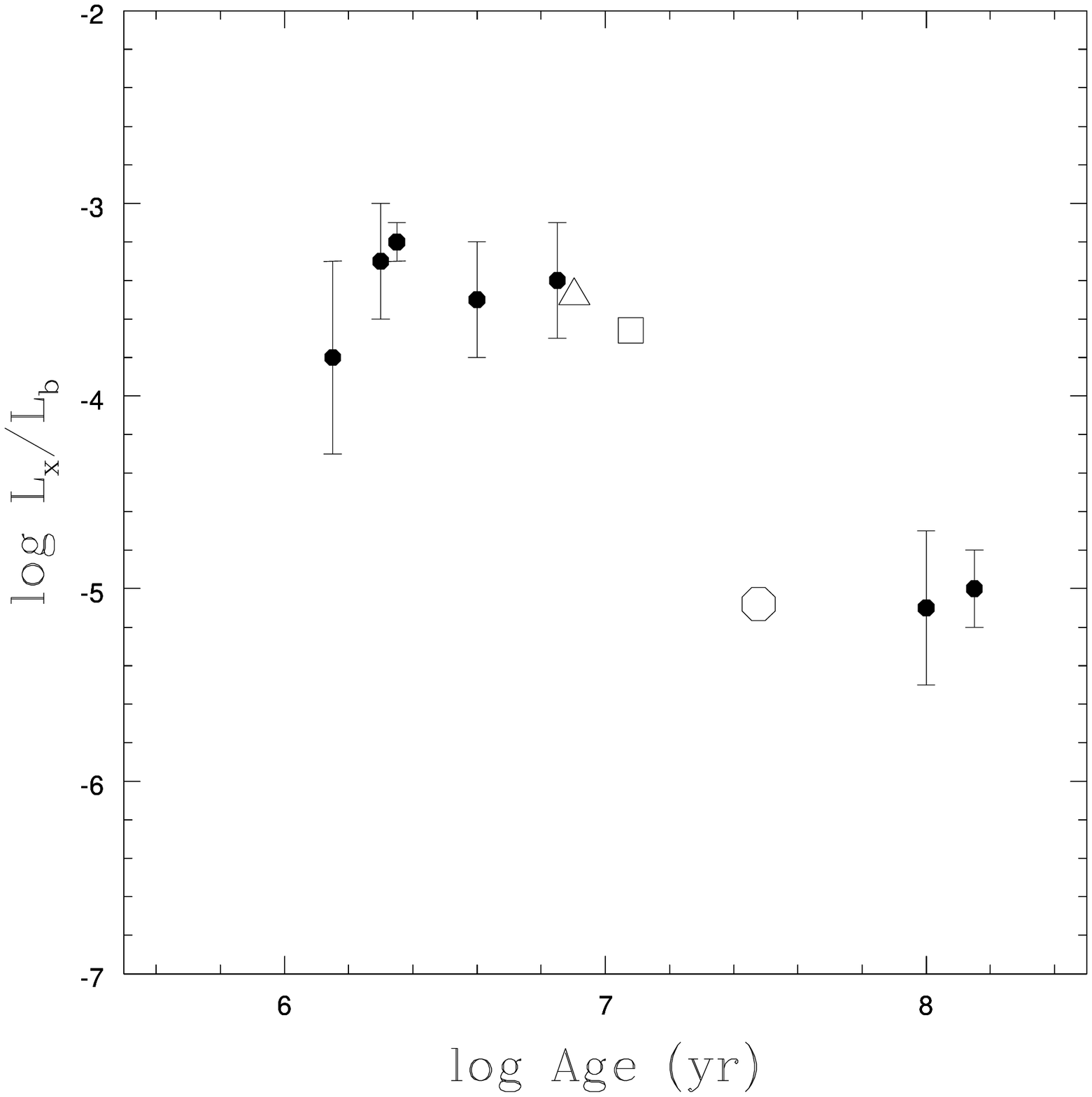}\\
\end{tabular}
\caption{As in Figure \ref{fig7} but for the high rotating mode.} 
\label{fig8}
\end{center}
\end{figure}
\end{center}

\begin{center}
\begin{figure}
\begin{center}
\includegraphics[angle=0,width=12cm]{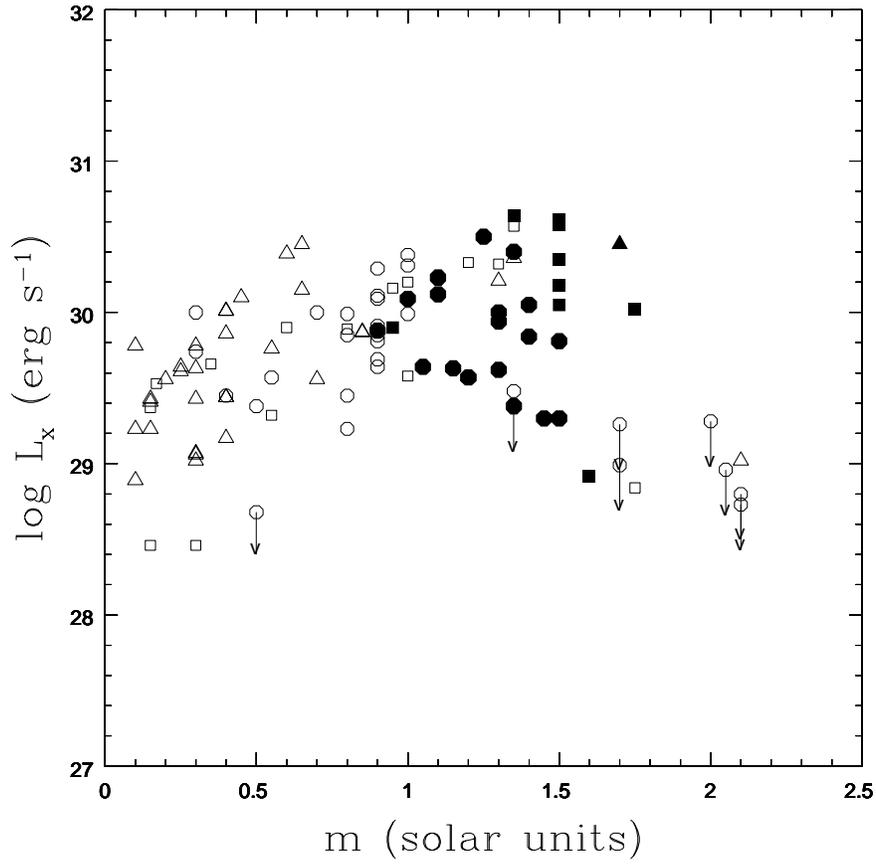}
\caption{$L_{x}$ \textit{vs} stellar masses for the associations. The F and G stars are represented by black symbols. Upper limit values are represented with arrows.} 
\label{fig9}
\end{center}
\end{figure}
\end{center}

\begin{center}
\begin{figure}
\begin{center}
\includegraphics[angle=0,width=12cm]{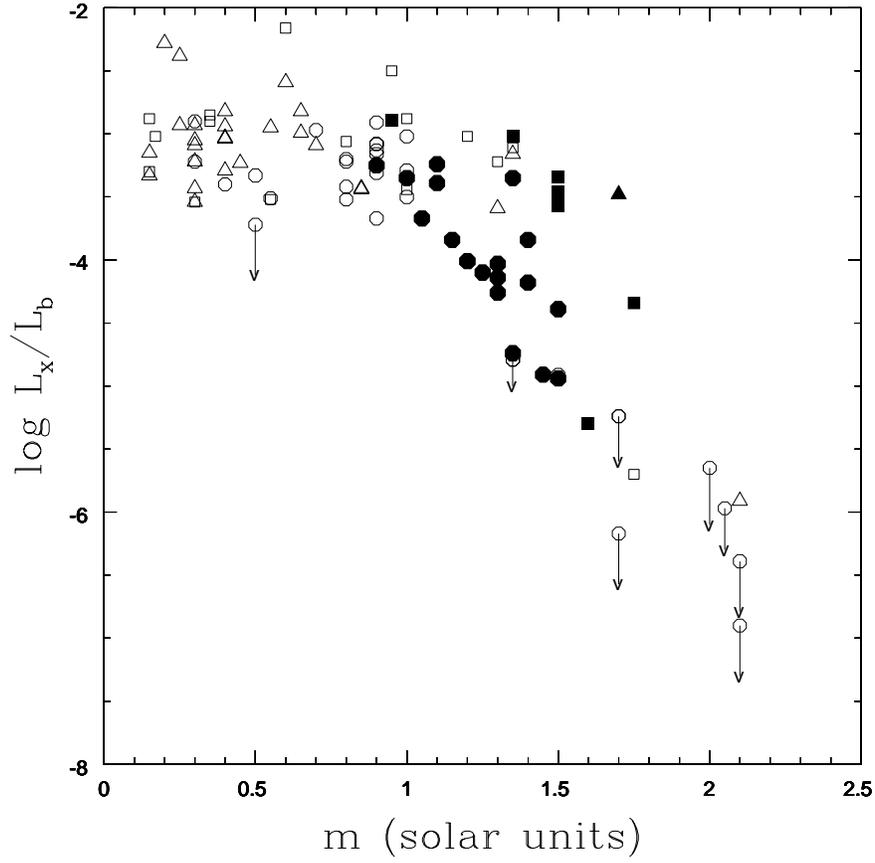}
\caption{$L_{x}/L_{b}$ \textit{vs} stellar masses for the associations. The symbols are the same as in Figure \ref{fig9}.} 
\label{fig10}
\end{center}
\end{figure}
\end{center}

Concerning the behavior of the PTTS of associations in Figures \ref{fig7} and \ref{fig8},
we can see that the $L_{x}$ indicator,
except for Tuc/HorA, does not present clear changes for the LRM and HRM.
Variations are more clearly seen by means of the $L_{x}/L_{b}$ indicator. First, we note that, different from the case of ONC, the LRM in TWA and BPMG present $L_{x}/L_{b}$
at the canonical saturated value. Apparently this could be due to the absence
of absorption accreting disks in these stars (see later). Nevertheless, as in
ONC, we can see the presence of a small desaturation tendency for HRM in
BPMG, also in an opposite direction to the linear MS relation. We note
nevertheless, that HRM in TWA depends on only one G5 (TWA19A)
star. By desaturation we mean a progressively decreasing of the X-ray indicators
with respect to the saturated values. We also note that this happens first in
the HRM which implies high masses and large $v_{0}(eq)$ values. The
desaturation effect appears clearly later at $\sim$30 Myr, indicated by both X-rays indicators,  only for
Tuc/HorA. In Figure \ref{fig7} and \ref{fig8} we show
the variations of these indicators only as a function of age for the two modes.
We have also considered in these figures the values presented by
Flaccomio et al.(2003b) for TTS with ages between 1 - 7 Myr and MS stars with
ages up to 140 Myr. Our data fills well a large part of the gap in the
Flaccomio et al. data. Here, we discuss more in detail this desaturation effect.
A discussion of the dependence of the X-rays properties on mass can be instructive.
In Figure \ref{fig9} we present $L_{x}$ versus stellar masses for the three
associations showing a similar increasing trend as observed in ONC (see references
mentioned above).
In Figure \ref{fig10} are shown the values of $L_{x}/L_{b}$ as a function of stellar
masses also for the three associations. A relative similar figure can be found
in
Song et al. (2003,Fig 19) and Song et al. (2004,Fig.1), although in these figures,
differently
than in our Figure \ref{fig9}, only new proposed members of TWA, BPMG
and Tuc/HorA (also included in this work) and colors (B-V) instead of mass are included. From Figure \ref{fig9} it can be seen that, even if TWA members are concentrated in the region
of very low masses, TWA and BPMG have, in general, similar $L_{x}/L_{b}$ values
independent of mass. A notorious difference appears for the older Tuc/HorA,
where stars with masses larger than $\sim$1.1$M_{\odot}$ (corresponding to stars of types
earlier than $\sim$G3) show a steep decline of the $L_{x}/L_{b}$ values. This
softening  X-rays branch is formed by G, F and A stars. Even if the observed F
stars are more numerous in Tuc/HorA than in BPMG, F stars in BPMG have larger
values of $L_{x}/L_{b}$ than the corresponding stars in Tuc/HorA. An age effect is
then into action as far as larger masses than 1.1$M_{\odot}$ is concerned. We must note that increasing masses (and also increasing stellar temperatures) are a measure of
the diminishing of the corresponding stellar convecting layers (see for
instance Pinsonneault, DePoy \& Coffee 2001). Now, what about rotation ?. Examining
the descending branch we can see that its beginning (see Table 3) at $log$ $(L_{x}/L_{b})$  $\sim$ -4.0, is formed by stars of types $\sim$G3 to F8 with $vsini$ values
around
$\sim$30 $km s^{-1}$, whereas at the faint values of $\sim$ -5.0 or less, all stars correspond to very
high rotators with $vsini$ values larger than 100 $km  s^{-1}$. From this, we can infer that
at this stage rotation is no more important. It is interesting to note that
this desaturation effect at $\sim$30 myr is due to the fact that at this age, which
nicely coincides with the age of Tuc/HorA, began the
stabilization of the convecting layers, that is, when for these masses they attain for the
first time, the minimun configuration (see for instance Cameron, Campbell \& Quaintrell (1995); Keppens, MacGregor \& Charbonneau (1995); Siess \& Livio (1997)). The convecting
layers are then so thin that an increase of rotation is unable to
maintain high values of $L_{x}/L_{b}$ resulting from a dynamo process.

We note however,in Figure \ref{fig10} that the two hotter stars of BPMG, with types F2
($vsini$=155 $km s^{-1}$) and one A7 ($vsini$ = 134 $km  s^{-1}$) present very low values of
$L_{x}/L_{b}$, apparently contradicting our mentioned age effect (this also applies to
the sole A star of TWA). From this, we conclude that the age effect for
the
desaturation process is only sensitive for early G and late F stars. Earlier F
and A stars will always show low $L_{x}/L_{b}$ values independent of age.

We note that in effect, a discussion based on the mass dependence of the X-ray indicators (as in Figure \ref{fig10}) can give a better insight on the physical origins of the X-ray softening than that based on the rotation period (Figure \ref{fig6}). In fact, a discussion based only in Figure \ref{fig6} would eventually lead us to conclude that the high rotation is the cause of the X-rays softening, which this is not the cause as seen above.

This kind of
behavior is in agreement with other recent studies based on the hardness
ratio HR1 used as a measure of the $L_{x}$ radiation (Kastner et al. 2003;
Suchkov, Makarov \& Voges 2003). Kastner et al. studied the same PTTS
associations considered
in this work and detected a clear decline of HR1 values for G, F stars of
Tuc/HorA relative to BPMG. They also show that HR1 values of the three
associations are somewhat smaller than those of younger TTS (see also
the predictions made about this in Suchov, Makarov \& Voges  2003). Suchov et al.
realized an extensive general analysis of X-ray radiation of F type stars
measured by Hipparcos. They found that a small group of extreme X-rays
radiators ($log L_{x} > 30.4$) is dominated by young PMS stars. In our case, we
found that almost 3 F stars are included in this range in BPMG, whereas only one in
Tuc/HorA. In TWA we found that other type of stars, present these extreme
larger $L_{x}$ values as is the case of the K7 TW Hya star.

In conclusion, it appears that the softening of X-ray radiation with age is
related to intrinsic stars properties and not to others effects like absorption by surrounding disks as discussed by Kastner et al. (2003). In
fact, with the exception of very few stars in TWA, which have classical TTS
fine
dusty disks, probably in their last stages, as TW Hya  (even if this star is
observed pole-
on)
and Hen600A (TWA3A), no more discs of this kind are seen in these
associations. Note also, that the protoplanetary disks detected around TWA 11A
and around the Beta Pictoris star are of a different nature than those of TTS.
On the other hand, recents surveys searching for the presence of warm dust
in members of TWA (Weinberger et al.2004) and in BPMG (Weinberger et al.2003)
have shown apart some few stars, a general absence of such warm dust.
No similar survey has been made around Tuc/HorA stars, but the known absence
of IRAS sources among their members, and considering that IRAS radiation is correlated to the presence of warm dust (Weinberger et al.2004), we can expect a similar absence of fine warm dust in Tuc/HorA.

At this stage we mention that the same behavior of $L_{x}/L_{b}$ versus temperature or mass, indicating the existence of a general desaturation effect at this age, is also found in young stellar clusters as IC2602, IC2391 with ages
of 30 Myr (Stauffer et al.1997) and NGC 2547 (35 - 54 Myr, Oliveira et al.2003). A similar
behavior as our Figure \ref{fig10} for these three mentioned clusters can be seen in Figure 1 of Jeffries, Totten \& James 2000.

How this saturation effect will continue to evolve in direction to the early MS stage? In Figures \ref{fig7} and \ref{fig8}, we show that the HRM attain the early MS $L_{x}/L_{b}$ values at 30 Myr whereas this is not yet the case for the LRM. ROSAT studies of the Pleiades ($\sim$100 Myr) member stars present a different behavior, where low mass stars with $vsini$ values smaller than 10 $km s^{-1}$ have unsaturated $L_{x}/L_{b}$ values, whereas stars with larger $vsini$ values exhibit almost constant saturated $L_{x}/L_{b}$ values. We believe that studies of other stellar systems with intermediate ages between 30 and $\sim$100 Myr will be helpful to understand these differences, especially considering that angular momentum will not probably be conserved anymore. In general, it is important that rotation periods for individual members of the PTTS associations be observed to confirm our results and that future comparisons with older stars clusters be made as function of rotation $v_{0}(eq)$ velocities in a mass to mass way.

\section{Summary and Conclusions}

We have chosen three nearby post-T Tauri stars associations with ages between 8
and 30 Myr (TWA, 8 Myr; BPMG, 12 Myr and Tuc/HorA, 30 Myr) to study the stellar
rotations and X-ray related properties. For all probable and possible published
members of these associations, we have estimated their model dependent stellar
masses and radii by means of their observed $(V-I_{c})$ colors. We collected all best
observed projected rotational velocities $vsini$ values. When possible, the X-rays
indicators $L_{x}$ and $L_{x}/L_{b}$ were taken from the literature. When this was not the
case, we calculated the fluxes $F_{x}$ and $F_{b}$ by standard methods. All distances
used to calculate some luminosities were taken from the literature.

From the distribution of stellar masses versus vsini values we chose \textit{a-priori},
two groups of masses: low mass stars between 0.1 and 1.4$M_{\odot}$ and high masses,
between 1.5 and 2.6$M_{\odot}$. This initial mass division, later justified by a
statistical analysis, is fundamental for all the conclusions of this work. Two
low and high rotational modes were then established. This first general study of rotation of post-T Tauri stars in associations lead us to the following conclusions:

Because no measurements of the stellar rotation periods exist for these stars,
we inferred, by means of a numerical approach, representative equatorial rotation
velocities obtained from the $vsini$ observed distribution for each mode and for
each association. A spin up was found for the high rotation mode in the whole
interval between 8 and 30 Myr which is near the maximum of the spin up found at 40 Myr in pre main sequence stellar models. The spin up is not present in the low mode.
In any case, by estimating a mean radius for each association, which in
general decreases as $t ^{-1/5}$, we found that a mean specific angular momentum is practically conserved during this interval of time for the high mass stars. However, this is not the case for the low mass stars where a decrease of the specific angular momentum by a mean factor two is present. Somehow, these low mass stars do not reach large surface velocities, as is the case of the high mass stars in order to conserve their specific angular momentum. 

By admitting that the star's mass distribution in the associations follows an
universal initial mass function, we estimated the star's members that remain to
be detected for different mass intervals. Simulating $vsini$ values for these
undetected stars, we showed that their distribution, together with the observed
$vsini$ values, is in agreement with a general spin up picture. Another result,
even more compatible with a spin up of the high rotating mode, is obtained by
means of a two dimensional (mass and observed vsini) K-S statistical analysis.

Taking into account the two X-rays indicators; $L_{x}$ and $L_{x}/L_{b}$, we found that
stars, specially in the younger TWA and BPMG associations, present some similar
properties to those of T Tauri stars in the Orion Nebula cluster (1 Myr). These
are: a direct $L_{x}$ correlation with mass and a slight dependence with rotation in
the sense of falling $L_{x}/L_{b}$  for lower rotation periods. This trend is opposite
to the linear relation valid for low rotators in the main sequence. A strong
desaturation effect appears for the high mode in Tuc/HorA. The nature of this
effect is better understood when $L_{x}/L_{b}$ is plotted against the stellar masses.
This desaturation at 30 Myr is sensitive only for high mass early G and late
type F stars. This effect appears to be due only to intrinsic stellar
properties. In fact, at $\sim$30 Myr (which coincides with the age of Tuc/HorA) the
stellar convective layers of these high mass stars, attain for the first time
their minimum configuration. Because of this, their even very large rotation
velocities appear unable to maintain high $L_{x}/L_{b}$ values. A similar behavior of
$L_{x}/L_{b}$ is also found, at the same age, in young clusters as IC2602,IC2391 and
NGC2547.

Our results fill the gap of the general variation of the X-rays indicators
between T tauri stars and early main sequence stars. Whereas in the low
rotation mode the transition appears to be soft, in the high rotation
mode the indicators reach, at 30 Myr, nearly the early main sequence values.

\acknowledgments

We are specially thankful to the anonymous referee for important suggestions that improved the general results of this work. We are also grateful to Vladimir G. Ortega for useful conversations and Dr I.Song for informations concerning X-ray sources of some new members of the associations.





\clearpage



\clearpage
\clearpage

\clearpage
\clearpage

\clearpage

\clearpage

\clearpage

\clearpage

\clearpage

\clearpage

\clearpage

\begin{deluxetable}{rrrrrrrrrrrr} 
\tablecolumns{12} 
\tablewidth{0pc}
\rotate
\tablecaption{TWA} 
\tablehead{ 

\colhead{Name} & \colhead{ST}   & \colhead{V}    & \colhead{B-V} & 
\colhead{V-$I_{c}$}  & \colhead{M}   & \colhead{R}  & \colhead{$vsini$} & \colhead{d}  & \colhead{log $L_{x}$} & \colhead{log $L_{x}/L_{bol}$} 
 & \colhead{Ref.}}
\startdata 

TWA Hya:TWA1	        &K7e	&10.92&	0.97&	1.70&	0.60&	1.0&	4	&56.4&	30.39&	-2.59 &1,7 \\
CD-$29^{\circ}$ 8887A:TWA2A&	M2e&	11.07&	1.48&	2.24&	0.30&	0.70&	13&	43&	29.43&	-3.54&1,7	\\
CD-$29^{\circ}$ 8887B:TWA2B&	M2&	... &	...&	2.80&	0.15&	0.6&	...&	43&	29.43&	...&1,2,7\\
Hen 3-600A:TWA3A&	M3e&	12.04&	1.49&	2.94&	0.15&	0.55&	20&	35&	29.23&	-3.33&1,7\\
Hen 3-600B:TWA3B&	M3.5&	13.70&	...&	3.60&	$\le$0.1&	$\le$0.5&	...&	35&	29.23&	...&1,2,7\\
HD 98800A:TWA4A&	K4/K5&	9.41&	1.15&	1.51&	0.85&	1.15&	10&	46.7&	29.87&	-3.43&1,7\\
HD 98800B:TWA4B&	K7+M1&	9.94&	1.28&	...&	0.85&	0.70&	...&	46.7&	29.87&	-3.44&1\\
CD-$33^{\circ}$ 7795A:TWA5A&	M1.5&	11.54&	1.48&	2.62&	0.30&	0.70&	36&	48&	29.78&	-3.05&1,2\\
CD-$33^{\circ}$ 7795B:TWA5B&	M8.5&	20.40&	...&	4.60&	$\le$0.1&	$\le$0.5&	...&	48&	29.78	&...&1,2\\
TYC 7183- 1477- 1:TWA6&	K7&	11.62&	1.31&	1.68&	0.65	&1.05&	55&	77&	30.15&	-2.82&1,2,7\\
TYC 7190- 2111- 1:TWA7&	M1&	11.65&	1.46&	2.44&	0.25&	0.70&	2&	38&	29.64&	-2.38&1,2,7\\
TWA 8A&	        M2&	13.30&	1.46&	2.41&	0.25&	0.70&	2&	21&	...&	...&1,2,7\\
TWA 8B&         	&	...&	&	&	&	&	5&	21&	...&	...&1,2\\
CD-$36^{\circ}$ 7429A:TWA9A&	K5&	11.26&	1.26&	1.62&	0.70&	1.00&	10&	50.3&	29.56&	-3.09&1,2,7\\
CD-$36^{\circ}$ 7429B:TWA9B&	M1&	14.00&	1.42&	2.58&	0.20&	0.60&	4&	50.3	&29.56&	-2.28&1,2,7\\
GSC 07766-00743:TWA10&	M2.5&	12.96&	1.43&	2.47&	0.25&	0.80&	2&	57&	29.61&	-2.93&1,2,7\\
HR4796A:TWA11A&	        A0&	5.78&	0.00&	&	2.10&	1.65&	152&	67.1&	29.02&	-5.91 & 1,10\\
HR4796B:TWA11B&	        M2.5&	13.30&	...&	&	0.30&	0.75	&...&	67.1&	29.02&	-3.22&1,2\\
RX J1121.1-3845:TWA12&	M2&	12,85?&	1.50&	2.25&	0.30&	0.75&	15&	32&	29.06&	-2.93&1,2\\
RX J1121.3-3447N:TWA13A&	M1e&	11.46&	1.44&	1.97	&0.40&	0.80&	10&	38&	29.44&	-3.03&1,7\\
RX J1121.3-3447S:TWA13B&	M2e&	12.40&	1.44&	1.97&	0.40&	0.80&	10&	38&	29.44&	-3.04&1,7\\
1RXS J111325.1-452344:TWA14&	M0&	11.85&	...&	2.85&	0.15&	0.60&	...&	46&	29.41&	-3.15&1,2\\
1RXS J123420.1-481514:TWA15A&	M1.15&	13.51&	...&	2.16&	0.40&	0.80&	22&	119&	30.01&	-2.82&1,2\\
1RXS J123420.1-481514:TWA15B&	M2&	14.00&	...&	2.19&	0.40&	0.80&	30&	119&	30.01&	...&1,2\\
1RXS J123456.1-453808:TWA16&	M1.5&	11.64&	...&	2.13&	0.30&	0.80&	11&	66&	29.63&	-3.43&1,2\\
1RXS J132046.5-461139:TWA17&	K5&	11.69&	...&	1.92&	0.45&	0.90&	45&	133&	30.10&	-3.23&1,2\\
1RXS J132137.0-442133:TWA18&	M0.5&	12.08&	...&	1.98&	0.40&	0.80&	20&	98&	29.86	&-3.29&1,2\\
HD102458:TWA19A&	        G5&	9.14&	0.70&	0.70&	1.70&	2.40&	25&	104.0&	30.45&	-3.48&1,2\\
HD102458B:TWA19B&	K7&	11.06&	...&	1.69&	0.65&	1.05&	10&	104.0&	30.45&	-2.99&1,2\\
A2-146:TWA20&	                M2&	13.40&	...&	2.30&	0.30&	0.70&	30&	50 g&	29.07&	-3.09 &2\\
HD298936:TWA21&	        M1&	9.79&	1.00&	...&	1.30&	1.50&	...&	69&30.21&-3.59&3\\
SSS 101726.7-535428:TWA22&	M5&	...&	1.80&	...&	$\le$0.1&	0.30&	...&	22&	28.89&	...&3\\
SSS 120727.4-324700:TWA23&	M1&	12.67&	1.48&	...&	0.40&	0.85&	...&	37&	29.17&	-2.94&3\\
GSC 08644-00802:TWA24&	        K3&	10.26&	0.95&	...&	1.35&	1.55&	...&	95&	30.36&	-3.16	&3\\
GSC 07760-00283:TWA25&	        M0&	11.36&	1.40&	...&	0.55&	0.95&	...&	44&	29.76&	-2.95	&3\\

\enddata 
\label{tbl-TWA}
\end{deluxetable}

\begin{deluxetable}{rrrrrrrrrrrr} 
\tablecolumns{12} 
\tablewidth{0pc}
\rotate
\tablecaption{BPMG} 
\tablehead{ 

\colhead{Name} & \colhead{ST}   & \colhead{V}    & \colhead{B-V} & 
\colhead{V-$I_{c}$}  & \colhead{M}   & \colhead{R}  & \colhead{$vsini$} & \colhead{d}  & \colhead{log $L_{x}$} & \colhead{log $L_{x}/L_{bol}$}  & \colhead{Ref.}
       }
\startdata 

HR9 &	F2	&6.19&	...&	0.46&	1.60&	2.20&	155&	39.10&	28.92&	-5.30 &4 \\
55 Eri	&	F0&	5.22&	...&	0.34&	1.75&	1.80&	95&	29.80&	30.02&-4.34 &4\\
GJ 3305&	M0.5&	10.59&	1.90& &		$\le$0.1&	$\le$0.38&	...&	29.80&	30.23&	-2.49 &4\\
HIP 23309&	M0/1&	10.01&	...&	1.79&	0.55&	0.90&	11&	26.30&	29.32&	-3.52 &4\\
HD 35850&	F7&	6.30&	...&	0.62&	1.50&	2.20&	50&	26.80&	30.35&	-3.52 &4\\
BetaPic&	A3&	3.85&	...&	0.20&	1.75&	1.70&	139&	19.30&	...&	... &4\\
AO Men&	        K6/7&	9.77&	...&	1.34&	1.00&	1.25&	13&	38.50&	30.20&	-2.88 &4\\
V343Nor&	K0&	8.14&	...&	0.93&	1.30&	1.70&	11&	39.80&	30.32&	-3.22 &4\\
V343NorB&	...&	14.80&	...&	2.90&	0.15&	0.45&	...&	39.80&	...&	... &3\\
HR6070&	A0&	4.80&	...&	1.01&	2.15&	1.70&	30&	43.00&	...&	... &4\\
HD155555A&	G5&	7.21&	...&	0.81&	1.35&	1.80&	37&	31.40&	30.64&	-3.02 &4\\
HD155555B&	K0&	8.08&	...&	1.09&	1.20&	1.30&	34&	31.40&	30.33&	-3.02 &4\\
HD155555C&	M4.5&	12.71	&...&	2.69&	0.17&	0.50&	...&	31.40&	29.53&	-3.02 &4\\
HD164249&	F5&	7.01&	...&	0.53&	1.50&	2.20&	...&	46.90&	30.58&	-3.48 &4\\
HR6749&	A5&	5.67&	...&	0.26&	1.80&	1.80&	...&	43.90&	...&	... &4\\
HD172555&	A7&	4.78&	...&	0.20&	1.75&	1.70&	134&	29.20&	28.84&	-5.70 &4\\
CD-$64^{\circ}$ 1208&	K7&	9.54&	...&	1.55&	0.80&	1.10&	150&	29.20&	29.89&	-3.06 &4\\
PZ Tel&	K0&	8.25&	...&	0.85&	1.35&	1.85&	63&	49.70&	30.57&	-3.11 &4\\
HR 7329&	A0&	5.05&	...&	0.04&	2.10&	1.65&	150&	47.70&	...&	... &4\\
HR7329B&	M7/8&	11.93&	...&	...&	...&	...&	...&	47.70&	...&	... &4\\
HD181327&	F5.5&	7.03&	...&	0.53&	1.50&	1.70&	16&	50.60&	...&	... &4\\
GJ799&	M4.5e&	11.02&	...&	2.90&	0.15&	0.70&	...&	10.20&	29.37&	-2.88 &4\\
GJ803	&M1e&	8.81&	...&	2.10&	0.35&	0.75&	9&	9.90&	29.66&	-2.85 &4\\
HD199143&	F8&	7.34&	...&	0.62&	1.50&	2.20&	120&	47.70&	30.61&	-3.34 &4\\
BD-$17^{\circ}$ 6128&	K7/M0&	10.60&	...&	1.30&	1.00&	1.25&	12&	47.70&	29.58&	-3.44 &4\\
HD 14082 B&	G2&	7.75&	0.62&	...&	1.50&	1.30&	...&	34.00&	30.05&	-3.46 &3\\
HD 14082&	F8&	6.99&	0.52&	...&	1.50&	1.10&	...&	39.40&	30.18&-3.57&3\\
BD+30 397A&	G9/K0&	10.12&	1.21&	...&	0.95&	1.75&	...&	42.30&	29.90&	-2.89&3\\
BD+30 397B&	K5/K6&	12.44&	1.40&	...&	0.60&	1.70&	...&	42.30&	29.90&	-2.16&3\\
BD+05 378&	G9/K0&	10.37&	1.22&	...&	0.95&	1.75&	...&	40.50&	30.16&	-2.50&3\\
GJ 3322&	M1/M2&	11.50&	1.50&	...&	0.35&	1.65&	...&	32.10&	29.60&	-2.90&3\\
GJ 871.1A&	M2&	12.16&	1.51&	...&	0.30&	1.65&	...&	23.60&	28.46&	-3.54&3\\
GJ 871.1B&	M3/M4&	13.43&	1.58&	...&	0.15&	1.65&	...&	23.60&	28.46&	-3.30&3\\

\enddata 
\label{tbl-BPMG}
\end{deluxetable}

\begin{deluxetable}{rrrrrrrrrrrr} 
\tablecolumns{12} 
\tablewidth{0pc}
\rotate
\tablecaption{Tuc/HorA} 
\tablehead{ 

\colhead{Name} & \colhead{ST}   & \colhead{V}    & \colhead{B-V} & 
\colhead{V-$I_{c}$}  & \colhead{M}   & \colhead{R}  & \colhead{$vsini$} & \colhead{d}  & \colhead{log $L_{x}$} & \colhead{log $L_{x}/L_{bol}$} & \colhead{Ref.}
       }
\startdata 

HD1466& 	F8& 	7.45	& 0.54& 	0.59& 	1.3& 	1.25& 	18& 	40.9& 	29.62& 	-4.14 & 5,6\\
HIP1910	& M0& 	11.47& 	1.39& 	1.95& 	0.4& 	0.50& 	18& 	46.3& 	29.45& 	-3.4& 5,6\\
HIP1993	& K7& 	11.47& 	1.35& 	1.81& 	0.5& 	0.60& 	17& 	37.4& 	$\le$28.68& 	$\le$-3.72& 5,6,8,9\\
HD2884& 	B9& 	4.38& 	-0.07& 	0.00& 	2.1& 	1.65& 	107& 	42.8& 	$\le$28.80& $\le$-6.39& 5,6,8,9\\
HD2885& 	A2+A7	& 4.55& 	0.15?& 	0.25& 	1.7& 	1.50& 	6& 	52.8& 	$\le$28.99	& $\le$-6.17& 5,6,8,9\\
HD3003& 	A0& 	5.06& 	0.04& 	0.00& 	2.1& 	1.65& 	78& 	46.5& 	$\le$28.73& 	$\le$-6.90& 5,6,8,9\\
HD3221& 	K4& 	9.63& 	1.05& 	1.38& 	0.8& 	0.80& 	110& 	45.9& 	29.99& 	-3.2& 5,6\\
CPD-64120& 	K1e& 	10.29& 	0.86& 	1.01& 	0.9& 	0.95& 	32& 	59& 	30.09& 	-2.91& 7\\
HD8558& 	G6& 	8.5& 	0.68	& 0.77	& 1.1& 	1.05& 	15& 	49& 	30.12& 	-3.39& 7\\
HD9054& 	K2& 	9.07& 	0.91& 	1.01& 	0.9& 	0.95& 	6& 	37& 	29.81& 	-3.31& 7\\
GSC8047-0232& 	K3& 	10.87& 	0.95& 	1.08& 	0.9& 	0.90& 	19& 	89& 	30.11& 	-3.08& 7\\
HD12039	&G4/G5& 	8.07& 	0.65& 	...& 	1.2& 	1.15& 	...& 	42.4& 	29.57& 	-4.01& 3\\
CD-53386& 	K3e& 	11.02& 	0.96& 	1.12& 	0.9& 	0.90	& 19& 	117& 	30.29& 	-3.08& 7\\
HD12894& 	F4& 	6.43& 	0.36& 	0.43& 	1.5& 	1.45& 	110& 	47& 	29.3& 	-4.94& 7\\
HD13183	& G5& 	8.63& 	0.65& 	0.76& 	1.1& 	1.05& 	23& 	50& 	30.23& 	-3.24& 7\\
CD-60416& 	K5& 	10.68& 	1.16& 	1.43& 	0.8& 	0.80& 	12& 	45	& 29.85& 	-3.22& 7\\
HD13246& 	F7& 	7.5& 	0.52& 	0.6& 	1.3& 	1.25& 	35& 	45& 	30.00& 	-4.26& 7\\
GSC8056-0482& 	M3e& 	12.11& 	1.48& 	2.33& 	0.3	& 0.40& 	37& 	42& 	29.74& 	-2.90& 7\\
GSC8491-1194& 	M3e& 	12.21& 	1.49& 	2.4& 	0.3& 	0.40& 	8& 	40& 	30.00& 	-3.22& 7\\
CD-53544	& K6e& 	10.21& 	1.26& 	1.6& 	0.7& 	0.70& 	80& 	40& 	30.00& 	-2.97& 7\\
GSC8497-0995& 	K6e& 	10.97& 	1.23& 	1.48& 	0.8& 	0.75& 	5& 	48& 	29.23& 	-3.52& 7\\
GSC8862-0019& 	K4e& 	11.67& 	1.04& 	1.25	& 0.9& 	0.85& 	6& 	98& 	29.85& 	-3.13& 7\\
CD-65149& 	K2e& 	10.19& 	0.83& 	...& 	1.0& 	1.00& 	75& 	84& 	30.31&	-3.02& 7\\
GSC84999-0304&	M0e&	12.09&	1.25&	1.54&	0.8&	0.75&	11&	77&	29.45&	-3.42& 7\\
TYC 5882- 1169-1&	K3/K4&	10.17&	1.01&	...&	0.9&	0.95&	...&	51&	30.35&-3.36& 3\\
HD30051&	F5&	7.12&	0.41&	...	&1.5&	1.50&	...&	58.1&	29.81&	-4.39& 3\\
HD35114&	F7/F8	&7.39&	0.51&	...&	1.4&	1.40&	...&	45.7&	30.05&	-3.84& 3\\
TYC 7600- 516-1	&K1&	9.58&	0.84&	...&	0.85&	1.00&	...&	53&	29.98&-3.25& 3\\
TYC 7065- 0879-1	&K4/K5&	11.23&	1.09&	...&	0.9&	0.90&	...&	70&	29.69&	-3.25& 3\\
HD40216&	F7&	7.46&	0.49	&...&	1.4&	1.40&	...&	54.3&	29.84&	-4.18& 3\\
HD43989&	F9&	7.95&	0.54&	...&	1.35&	1.35&	...&	49.8&	30.40&	-3.35& 3\\
HD44627	&K1&	9.13&	0.86&	...&	1&	1.00&	...&	45.5&	29.99&	-3.29& 3\\
HD49855&	G3&	8.94&	0.7&	...&	1.15&	1.10&	...&	56.5&	29.63&	-3.84& 3\\
HD55279&	G7&	10.11&	0.96&	...&	0.9&	1.00&	...&	64.1&	29.88&	-3.25& 3\\
HD174429&	K0&	8.25&	0.78&	0.85&	1&	1.00&	63&	49.6&	30.38&	-3.5& 5,6,8,9\\
HD177171&	F7&	5.16&	0.53&	0.63&	1.25&	1.20&	26&	52.4&	30.50&	-4.1& 5,6,8,9\\
HD181296&	A0&	5.05&	0.02&	...&	2.05&	1.65&	150&	47.6&	$\le$28.96&	$\le$-5.97& 5,6,8,9\\
HD181327&	F5&	7.03&	0.48&	0.53&	1.35&	1.35&	16&	50.6&	$\le$29.38	&$\le$-4.74& 5,6,8,9\\
HD191869S&	F6.5&	7.93&	0.49&	0.54&	?&	1.35&	33&	65.3&	29.48	&-4.79& 5,6,8,9\\
HD191869N&	...&	8.07&	...&	0.57&	1.35&	1.30&	30&	65.3&	29.48&	-4.79&5,6,8,9\\

HD200798&	A5&	6.69	&0.24	&0.28&	1.7&	1.50&	120&	66.4&	$\le$29.26&	$\le$-5.24&5,6,8,9\\
HD202917&	G5&	8.68&	0.69&	0.8&	1&	1.05&	13&	45.8&	30.09&	-3.35& 5,6,8,9\\
HD202947&	K0&	8.91&	0.85&	1.05&	0.9&	0.95&	13&	46&	29.91&	-3.16& 5,6,8,9\\
HIP107345&	M1&	11.62&	1.4&	1.83&	0.5&	0.60&	14	&42.3&	29.38&	-3.33& 5,6,8,9\\
HD207575&	F6&	7.22&	0.51&	0.56&	1.3&	1.30&	30&	45.1&	29.94&	-4.03& 5,6,8,9\\
HD207964&	F3&	5.9&	0.39&	0.46&	1.5&	1.45&	110&	46.5&	29.30&	-4.91& 5,6,8,9\\
PPM366328&	K0&	9.67&	0.8&	1&	0.55&	0.90&	...&	50&	29.57&	-3.51& 5,6,8,9\\
HD222259S&	G5/G8&	8.49&	0.78&	0.78&	1.05&	1.05&	16&	46.2&	29.64&	-3.67& 5,6,8,9\\
HD222259N&	...&	9.73&	1.14&	...&	0.9&	0.90&	13&	46.2&	29.64&	-3.67& 5,6,8,9\\
HD224392&	A1&	5.01&	0.06&	0.06&	2&	1.60&	152&	48.7&	$\le$29.28	&$\le$-5.65& 5,6,8,9\\

\enddata 

\label{tbl-TucHorA}

\tablecomments{Probable and possible members of TWA (Table 1), BPMG (Table 2) and Tuc/Hor A (Table 3). Columns from left to right are: stellar name, spectral type, visual magnitude, $B-V$, $V-I_{c}$, stellar mass in solar units, stellar radius in solar units, observed projected rotational velocity in $km$ $s^{-1}$, distance in $pc$ (values with decimal correspond to Hipparcos values), X-ray luminosity $L_{x}$ in $erg$ $s^{-1}$, ratio of $L_{x}/L_{b}$ where $L_{b}$ is the bolometric luminosity. References: (1) Torres et al.(2003), (2) Reid (2003), (3) Song, Zuckerman \& Bessell (2003,2004), (4) Zuckerman et al.(2001a), (5) Zuckerman \& Webb (2000), (6) Zuckerman, Song \& Webb (2001b), (7) Torres et al.(2000), (8) Stelzer \& Neuh$\ddot{a}$user (2000), (9) Stelzer \& Neuh$\ddot{a}$user (2001), (10) Royer et al.(2002)}

\end{deluxetable}

\clearpage

\begin{deluxetable}{rrrrrrrrrr} 
\tablecolumns{10} 
\tablewidth{0pc}
 \rotate
\tablecaption{}
\tablehead{ 

\colhead{Association} & Age &  \colhead{$v_{0}(eq)$}& \colhead{$log <P>$} & \colhead{$log <L_{x}>$} & \colhead{$log <L_{x}/L_{b}>$}  & \colhead{N}}

\startdata

TWA &8  & 28/113   & 0.16/-0.04& 29.85/30.45  &-2.93/-3.48 & 31/2  \\
BPMG &12 &  25/139  & 0.39/-0.19& 30.07/30.24  &-2.86/-3.66 & 19/13 \\
Tuc/HorA &30 &  32/159  &  0.19/-0.31& 29.97/29.28  &-3.37/-5.08 & 41/8\\

\hline
\hline

\colhead{Association} & \colhead{$<M>$}  & \colhead{$<R>$}& \colhead{$k^{2}_{conv}$} & \colhead{$k^{2}_{rad}$} & \colhead{$k^{2}_{total}$} & \colhead{$J/M$}\\

\hline

TWA & 0.42/1.90 &0.8/2.03 & 0.200/0.000 & 0.000/0.033 & 0.200/0.033 & 3.1$\times10^{16}$/5.3$\times10^{16}$ \\
BPMG & 0.67/1.68 &1.20/1.79 & 0.183/0.000 & 0.015/0.043 & 0.198/0.043 & 4.1$\times10^{16}$/7.4$\times10^{16}$   \\
Tuc/HorA & 0.96/1.77 &0.98/1.54 & 0.000/0.000 & 0.084/0.043 & 0.086/0.043 & 1.9$\times10^{16}$/7.3$\times10^{16}$ \\

\enddata

\label{tbl-med}

\tablecomments{Mean characteristic values for low rotation / high rotation modes for the three associations. Columns in order are: for the upper table: association name, age in Myr, equatorial velocity in $km$ $s^{-1}$, rotation period in days, X-ray luminosities in $erg$  $s^{-1}$, ratio of luminosities and total number of stars considered in each case. In the lower table: association name, stellar mass in solar units, stellar radius in solar units, square of the convective radius of gyration, square of the radiative radius of gyration, square of the total radius of gyration and specific angular momentum in $cm^2$ $s^{-1}$.}

\end{deluxetable}

\end{document}